\documentclass[aps,reprint,prb]{revtex4-2}

\usepackage{graphicx,comment}
\usepackage{dcolumn}
\usepackage{bm}
\usepackage{amsmath,amssymb,amsfonts}
\usepackage[english]{babel}
\usepackage{natbib}
\usepackage{braket}
\usepackage{float}
\usepackage[caption=false]{subfig}
\usepackage[dvipsnames]{xcolor}
\newcommand\tinyopenone{\vcenter{\hbox{\scalebox{0.5}{$\openone$}}}}

\begin{document}
\title{Bound states and controllable currents on Topological Insulator surfaces with extended magnetic defects}

\date{\today}

\author{Eklavya Thareja}
\email{ethare1@lsu.edu}
\affiliation{Department of Physics and Astronomy, Louisiana State University, Baton Rouge, LA 70803}

\author{Ilya Vekhter}
\email{vekhter@lsu.edu}
\affiliation{Department of Physics and Astronomy, Louisiana State University, Baton Rouge, LA 70803}

\begin{abstract}
We show that a magnetic line defect on the surface of a topological insulator generically supports two {distinct} branches of spin-polarized and current carrying one-dimensional bound states. We identify the components of magnetic scattering that lead to the bound states. The velocity, and hence spin texture, of each of those branches can be independently tuned by a magnetic field rotated in the plane of the surface. We compute the local net and spin-resolved density of states {as well as spin accumulation and charge currents.}
The net spin polarization and current due to both bound and scattering states vary stepwise as a function of the electrostatic and magnetic components of the scattering potential, and can be tuned by an applied field.  We discuss stability of the bound states with respect to impurity scattering.
\end{abstract}

\maketitle

\section{Introduction} Spin-momentum locking of the surface states in 3D topological insulators (TIs) protects them from backscattering except when the perturbing potential breaks time-reversal symmetry.~\cite{Zhang_RMP_2011,Kane_RMP_2010, Rev_hasan} {Common belief is that for nonmagnetic scattering the salient features of these states, such as the Dirac spectrum, remain intact.} However, resonance (nearly localized) states  which appear in the vicinity of individual impurities~\cite{Biswas2010,Black-Schaffer2012,Black-Schaffer2012a,Sablikov2015,Shiranzaei2017} have been observed in experiments~\cite{Alpichshev2012,Teague2012,Xu2017}. At finite impurity density, {for randomly distributed scattering centers, the entire low-energy part of} Dirac dispersion of the surface topological states {may be} modified due to  hybridization with the impurity {resonances}.~\cite{Miao2018,Clark2022}. Impurity signatures appear not only in the total density of states, but also in the spin textures  arising from the spin-momentum locking~\cite{Biswas2010}.

Multiple scattering on impurity clusters, {may almost lift}  the topological protection~\cite{Fransson2014}, generating gaplike features { for quasi-regularly arranged impurity centers}. These observations raise the question of whether spatially extended defects~\cite{Xu2017}  can be used to control spin textures, or spin and charge currents at topological surfaces. In this paper we show how this can be achieved in a minimal model of extended defects.

\begin{figure}[t]
	\centering
	\includegraphics[width=0.4\textwidth]{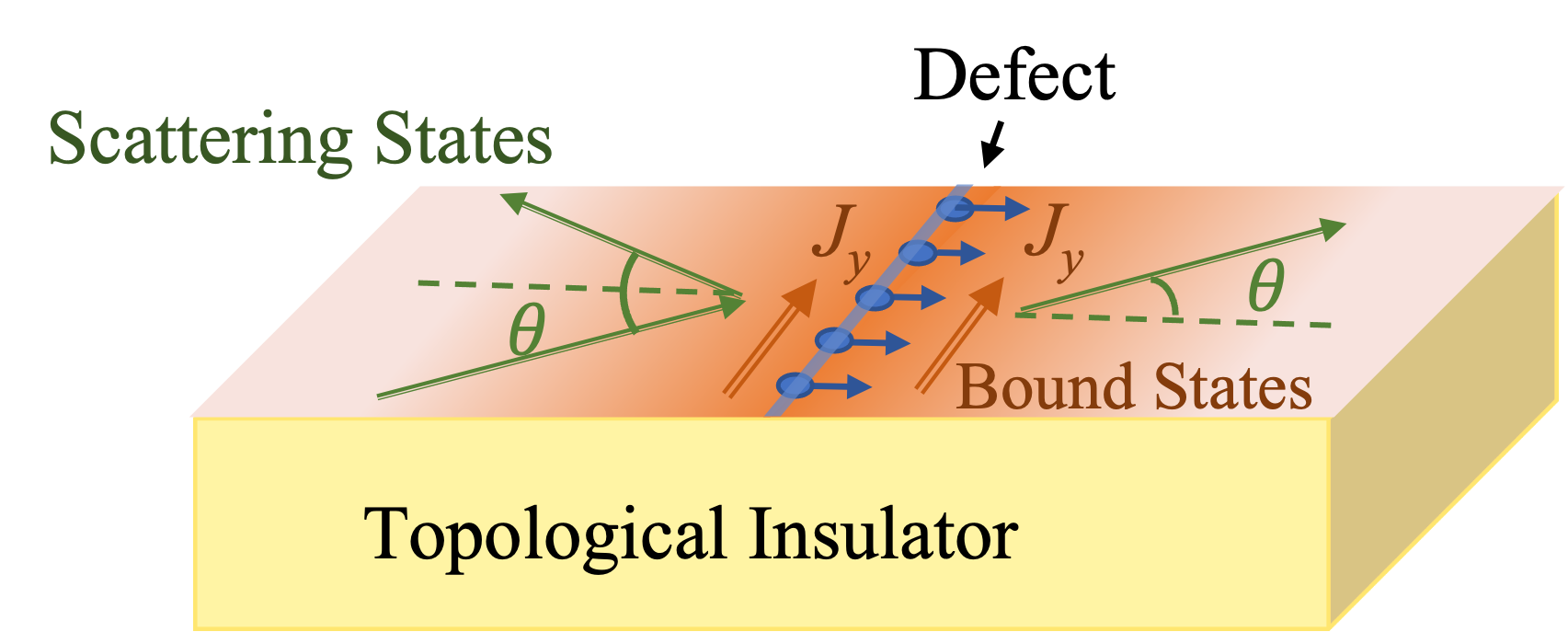}
	\caption{Topological Insulator with a line defect at the surface supports localized ({shaded} red) states in addition to scattering electrons. {Both in-plane and out-of-plane spin accumulation and charge currents parallel to the defect line result from magnetic scattering, and their magnitudes can be controlled via an external magnetic field.} }
	\label{ti_with_step}
\end{figure}

The simplest such defect is a line~\cite{Biswas2011,Sen2012,Liu2012,Brey2014a,Zhou2016}, realized experimentally near surface steps~\cite{Alpichshev2011, Fedotov2019}.  Both localized (1D states propagating along the line) and scattering states have been studied for electrostatic potential on a line or strip~\cite{Yokoyama2010,Biswas2011,Liu2012,Xu2018}, while scattering states were {also} investigated for a magnetic strip~\cite{Mondal2010}.  We consider the combined effect of electrostatic and magnetic scattering on a line defect, and compute the resultant spin textures and charge currents, sketched in Fig.~\ref{ti_with_step}. If magnetic scattering is due to the adsorbed atoms with classical magnetic moments we show that the in-plane magnetic field, that orients those moments, controls magnetization and net  charge current. The current varies stepwise with the field direction, with the values determined by the chemical potential position in the Dirac cone. These results are parametrically stable with respect to random point-like impurity scattering.
Our results open the possibility of using line defects at surfaces of 3D topological insulators to create current and magnetization channels.

\section{\label{sec:model}Model} We model the surface states by a Dirac Hamiltonian in spin space~\cite{Zhang_RMP_2011,Kane_RMP_2010}, and include a line defect with both magnetic and non-magnetic scattering,
\begin{equation}
	H = v(\bm \sigma\times \widehat{\bm k})_z + U_0\openone\delta(x) + \sum_{i=x,y,z}U_i\sigma_i \delta(x)\,.
	\label{full_hamiltonian}
\end{equation}
Here $U_0$ is the electrostatic potential, $U_i$'s describe magnetic scattering, $\sigma_i$ are the Pauli matrices in spin space~\cite{Liu2010}, and $\widehat{\bm k}$ is the momentum {operator}. The established agreement between the results from a 3D-based description of the surface states~\cite{Black-Schaffer2012,Sablikov2015,Shiranzaei2017,Xu2018} and the effective surface models~\cite{Biswas2010,Black-Schaffer2012a,Yokoyama2010} justifies this choice {of the Hamiltonian}. Since Eq.~\eqref{full_hamiltonian} is written in the long-wavelength approximation near the $\Gamma$ point for typical tetradymite topological insulators~\cite{Zhang_RMP_2011,Kane_RMP_2010, Rev_hasan}, the $\delta$-function approximation is valid for the potentials that decay on the scale $l_0\sim v/E_G$, where $E_G$ is the bulk energy gap.
The first term above yields helical linearly dispersing states in the absence of scattering. We assume ferromagnetic alignment of the spins at the defect line~\cite{Liu2009,Efimkin2014}, but allow for rotation of the moments by an external in-plane field, {thus changing the values of $U_i$'s}. 

It is instructive first to perform a symmetry analysis of Eq.~\eqref{full_hamiltonian}. The momentum along the defect, $k_y$, is a good quantum number, and can be used to classify the eigenstates. {In Eq.~\eqref{full_hamiltonian} the first term is both particle-hole and time-reversal symmetric.} $U_0$ breaks the particle-hole symmetry, while $U_i$ breaks time-reversal symmetry {of the Hamiltonian}. The latter allows spin accumulation and charge currents, but those are further constrained by symmetry. {When only magnetic scattering due to $U_x$ is present, {the} mirror symmetry about $x$-axis is broken while the mirror symmetry about the $y$-axis is intact. Thus, we expect the spin components $s_y$ and $s_z$ to change sign across the defect, while $s_x$ remains continuous. On the other hand, when only magnetic scattering due to $U_y$ is present, the mirror symmetry about the $y$-axis is broken while mirror symmetry about $x$-axis is intact.} This, combined with translational symmetry along $y$, implies that $s_y$ will be constant along $y$-axis, while $s_x$ and $s_z$ must vanish. {Later we will see that $s_y$ also vanishes} for this case.


\section{Boundary conditions.} The Hamiltonian has to be supplemented by the boundary conditions at the defect line. For the linear in momentum Dirac systems, the wave function is discontinuous across the boundary~\cite{McKellar1987,McKellar1987a}. The boundary conditions for scalar potentials { and step discontinuity} were investigated in graphene~\cite{McCann2004,Akhmerov2008,Basko2009c} and topological insulators~\cite{Sen2012,Enaldiev2015,Zhou2016}. { Similar boundary condition arise at an edge between two surfaces on different planes as elucidated in Ref.~\onlinecite{Brey2014a}.}

Direct integration of the eigenvalue equation, $H\psi = E\psi$ ~\cite{McCann2004,Akhmerov2008,Basko2009c}  gives $\psi(x) = e^{\int_{x_0}^x\hat{O}dx} \psi(x_0)$, where
\begin{equation}
v \hat{O}= -i\sigma_y\left[E-(U_0\openone+\bm {U\cdot \sigma})\delta(x)-v\sigma_xk_y\right]\,.
\end{equation}
Evaluating the integral across the defect line we find
\begin{equation}
	\psi(0^+) = e^{\frac{(U_0i\sigma_y + U_x \sigma_z + iU_y \tinyopenone - U_z\sigma_x)}{v}}\psi(0^-) {\equiv} \mathcal{M} \psi(0^-),
	\label{boundary_condition}
\end{equation}
{where matrix $\mathcal{M}$ encodes the boundary condition.} In Eq.~\eqref{boundary_condition}, $U_y$ appears as pure phase, and does not affect the observables, hence we set $U_y=0$. In contrast, $U_0$ rotates the spinor, while $U_x$ and $U_z$ also change the magnitude of the spinor components. This boundary condition enforces continuity of the $x$-component of the current, and hence satisfies particle conservation~\cite{Zhang2012,Enaldiev2015,Asmar2017,Alspaugh2022}. {Note that, while the general form of the matrix $\mathcal{M}$ could be inferred from the current conservation (in analogy with how it was derived for potential impurities in Ref.~\onlinecite{Alspaugh2022}), Eq.~\eqref{boundary_condition} gives the connection between the specific components of that matrix and corresponding scattering potentials. This is important for our subsequent analysis of the influence of the magnetic field, see Sec.~\ref{sec:control}.}
Below we set $U_z=0$ since a) dipolar interactions favor in-plane spin orientation; b) out-of-plane magnetic field opens a gap in the surface states spectrum removing low energy extended states; c) we verified the absence of bound states near the defect lines for  $U_z\neq 0$.

 \begin{figure}[t]
	\centering
	\includegraphics[width=0.4\textwidth]{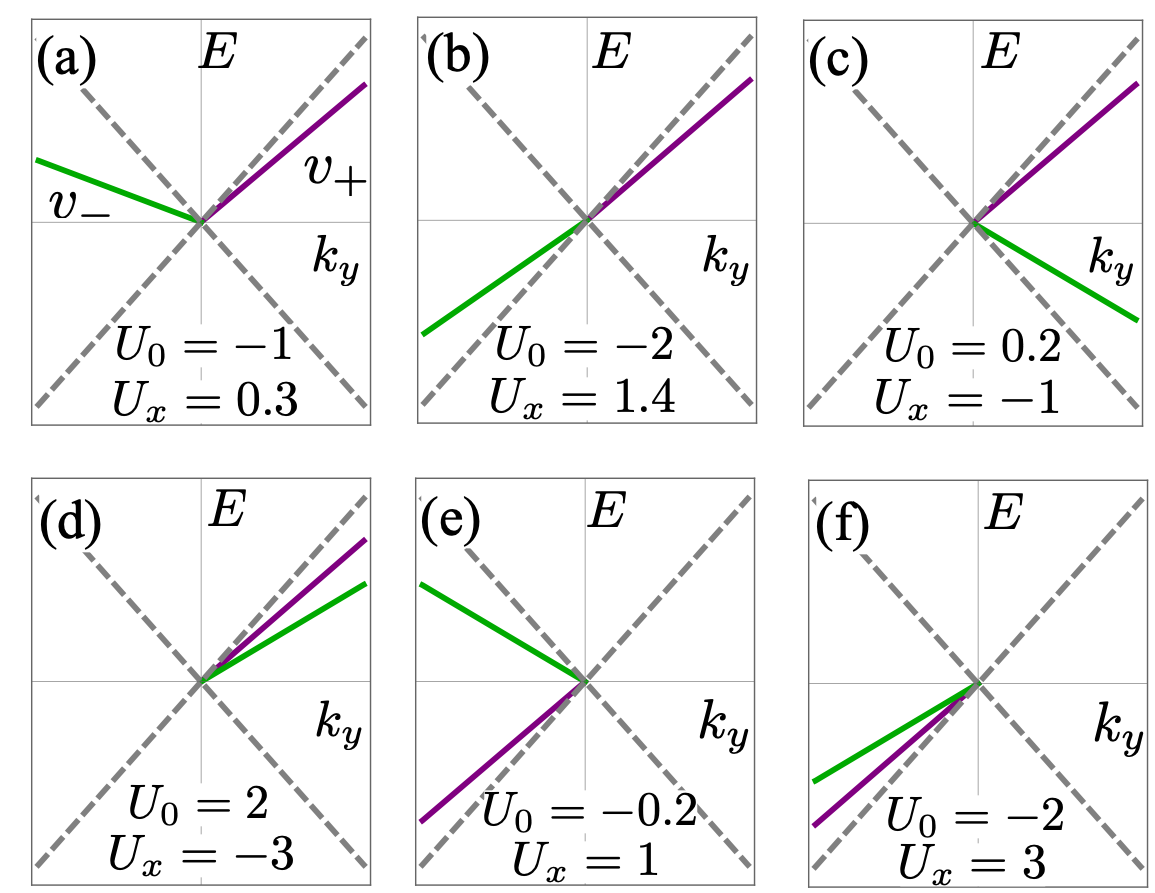}
	\caption{Energy spectrum of the bounds states (purple/green lines) for different values of $U_0$ and $U_x$. We set $v=1$, and denoted the Dirac cone with a dashed line. The velocities $v_\pm$, Eq.~\eqref{v_eff} , have different dependence on $U_0$ and $U_x$, see panels (a)-(d) and (e)-(f),{ and hence can be separately tuned.} {In addition to the six cases shown, there are six additional cases for a different set of values of the potentials} which yield bound states with the energies of the opposite sign, to those above, at each $k_y$, which we
refer to as cases (a')-(e') in Fig.~\ref{Charge_current_vary_Ux}.}
	\label{dispersion}
\end{figure}

\section{Bound states}
\label{sec:BS}

 For $U_x,U_0\neq 0$ the Hamiltonian in Eq.~\eqref{full_hamiltonian}, subject to the boundary conditions above, supports one-dimensional states bound to the defect of the form
{
\begin{equation}
\psi^{\pm}(x>0,k_y) =
\begin{pmatrix}
\sin\frac{\alpha_\pm}{2}\\
\pm\cos\frac{\alpha_\pm}{2}
\end{pmatrix} \sqrt{\frac{\lambda_\pm}{b}}e^{-\lambda_\pm x} e^{i k_y y},
\label{wavefn_bound_right}
\end{equation}
and

\begin{equation}
\psi^{\pm}(x<0,k_y) =
\begin{pmatrix}
\cos\frac{\alpha_\pm}{2}\\
\pm\sin\frac{\alpha_\pm}{2}
\end{pmatrix} \sqrt{\frac{\lambda_\pm}{b}}e^{\lambda_\pm x} e^{i k_y y}.
\label{wavefn_bound_left}
\end{equation}
where $b$ is normalization length along $y$ and $\lambda_\pm>0$ {is the inverse localization length.}
Imposing the boundary condition, {see Appendix A for details, gives}
\begin{equation}
\tan\frac{\alpha_\pm}{2} = \frac{\eta \cosh\eta + (U_x/v)\sinh\eta}{\eta\mp (U_0/v) \sinh\eta}\,,
\label{eq:alpha}
\end{equation}
{with} $\eta=\sqrt{U_x^2-U_0^2}/v$. In the same notation the dispersion, $E_\pm=\pm vk_y\sin\alpha_\pm\equiv v_\pm k_y$, with the effective velocity,
\begin{equation}\label{v_eff}
  \frac{v_\pm}{v}=- \left(\frac{U_x \mp U_0\cosh\eta}{U_0\mp U_x \cosh\eta}\right)\,.
\end{equation}
and the inverse localization length,
\begin{equation}
\lambda_\pm=k_y\cos\alpha_\pm = \pm k_y\frac{\eta \sinh\eta}{(U_0/v) \mp (U_x/v) \cosh\eta}.
\label{eq:lambda}
\end{equation}
For $|U_0|>|U_x|$, {the same equations hold if one takes} $\eta\rightarrow |\eta|$, and replaces hyperbolic functions by their trigonometric counterparts. }{Note that the localization length diverges ($\lambda_\pm\rightarrow 0$) at long wavelengths ($k_y\rightarrow 0$).}

The condition $\lambda_\pm >0$, {combined with Eq.~\eqref{eq:lambda},} means that the range of existence ($k_y>0$ or $k_y<0$) for each branch is determined by the sign of $\cos\alpha_\pm$, and hence depends on $U_x$ and $U_0$. In each case, the sign of $v_\pm$ determines whether the branch is above or below the Dirac point. {Several representative cases are shown in Fig.~\ref{dispersion}, and other arrangements of the bound state branches can be inferred from those, as discussed in the caption.}
Since $|v_\pm|\leq v$  the bound states are always ``outside'' the Dirac cone.

{From the above, the dimensionless parameter characterizing the strength of the scattering is $U_i/v$. Below we explore the entire range of the values for the scattering potentials, however, it is helpful to get a qualitative feel for the magnitudes involved. In Ref.~\onlinecite{Xu2017} for Bi$_2$Te$_3$ the experimental data for the scalar potential at the surface step were fit with the local line potential of  $V_0=3.8$ eV.
Assuming $V_0$ has the range comparable to the in-plane lattice constant , $a\sim 4.38$ \AA, we estimate $U_0\simeq V_0 a\approx 16.6$ eV$\cdot$\AA. The Dirac velocity in Bi$_2$Te$_3$ is $v\approx 4$ eV$\cdot$\AA, yielding $U_0/v\approx 4.1$. The same authors analyzed individual impurity resonances in Bi$_2$Te$_3$  for comparable values of the scalar and magnetic potentials ~\cite{Xu2017}, and we take that as an indication that a wide range of parameter values can be accessed experimentally. Of course, only comparison with detailed ab initio calculations can verify this in full, but such calculations are beyond the scope of our discussion here.}

When $U_x=0$ we recover the results of Refs.~\onlinecite{Biswas2011,Yokoyama2010}, and find two symmetric branches above or below the Dirac point, with 
$|v_\pm|=v\cos(U_0/v)$~\cite{Yokoyama2010,Levitov2015,*Levitov2018}.  Magnetic scattering breaks the symmetry between  $k_y$ and $-k_y$, selectively controlling the sign of $v_\pm$ and allowed signs of $k_y$ for each branch,
see Fig.~\ref{dispersion}(a)-(f). To the best of our knowledge, this behavior  has not been recognized previously.

When $\sqrt{U_0^2-U_x^2}/v=\pi n$, both branches merge with the Dirac cone and the bound state disappears \footnote{This condition can be shown to emerge naturally if one treats the $\delta$-function potential as a limiting case of a defect strip with finite width and considers interference of the reflected and transmitted waves at each boundary ~\cite{Eklavya1,Eklavya2}. }. For any $U_0$ there exists at least one value of $U_x$ where $v_+=0$ or $v_-=0$ \footnote{In this case interactions become important~\cite{Levitov2015,*Levitov2018}, and we leave this to a future discussion.}, generalizing the  condition $U_0=(n+1/2)\pi$~\cite{Levitov2015,*Levitov2018} for $U_x=0$.

The inverse localization length is $\lambda_\pm\simeq k_y \sqrt{1-(v_\pm/v)^2}$, and hence the states away from the Dirac point with a smaller velocity are better localized.  {For the bound states described by the spinors {in Eq.~\eqref{wavefn_bound_right} and~\eqref{wavefn_bound_left}},} the expectation value of the spin component $s_y=0$, while $s_x\propto\lambda_\pm v_\pm$. 
The $z$ component changes sign across the defect line, {i.e. has opposite signs for $x>0$ and $x<0$, as expected from our symmetry analysis above,} and we find {the magnitude} $s_z\propto\lambda^2_\pm/k_y$. 
Thus, flatter dispersion results in stronger out-of-plane polarization.  For $U_x=0$, the branches are symmetric, and 
hence only for the time-reversal broken states such as in Fig.~\ref{dispersion}(a)-(f) {and the corresponding complementary cases discussed in the caption} we observe a net polarization.

\section{Scattering states}
\label{sec:scattering}

\begin{figure}[t]
		\centering
		\includegraphics[width=0.6\columnwidth]{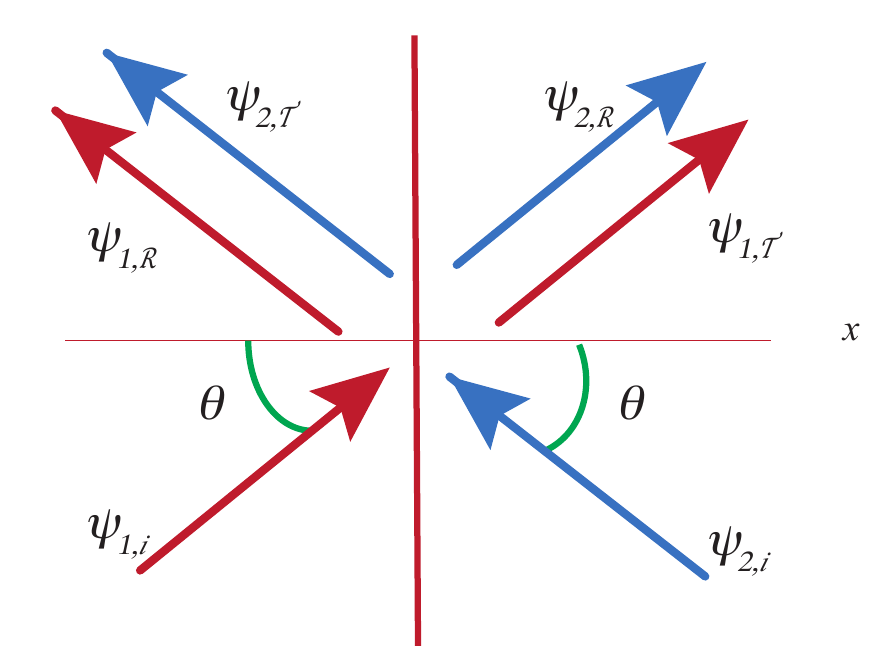}
		\caption{Scattering processes discussed in Sec.~\ref{sec:scattering}.}
		\label{fig:scattering}
	\end{figure}

{In addition to creating the bound states the defect also scatters the states in the Dirac continuum. The corresponding processes are shown in Fig.~\ref{fig:scattering}. The energy, $E(\bm k)=vk$, and the momentum along the defect, $k_y$, are conserved. Thus, the quasiparticle coming towards the defect at an angle $\theta$ with the positive $x$-axis, with the momentum $\bm k_i=k(\cos\theta,\sin\theta)$ has a reflected component with ${\bm k_\mathcal{R}} = (-k\cos\theta, k\sin\theta)$, in addition to the transmitted component with the same momentum $\bm k_i$. We label the corresponding wavefunctions by the subscript $1$ below.  It combines with the quasiparticle coming towards the defect from the opposite side, $x>0$, at an angle $\theta$ with the negative $x$-axis, where the incoming momentum is $\bm k^\prime_i=\bm k_{\mathcal R}$, with the wave functions labeled by superscript $2$.  Below we determine the transmission and reflection coefficients for these processes.}


 	
 {To do this we take into account that the wave functions of the helical quasiparticles have the spinor form $(i, he^{i\varphi})^T$, where $h = \mbox{sgn}(E)$ is the helicity, and $\varphi$ is the angle between the direction of its momentum and  the positive $x$-axis. For the incoming quasiparticles with momentum $\bm k_i$ combining the incident and reflected parts the of wavefunction in the region $x<0$ gives}
\begin{multline}
	\psi_1^h(x<0,k,\theta) \equiv\psi_{1,i}+\mathcal{R}\psi_{1,{\mathcal R}}
\\= \Xi\begin{pmatrix}
		i\\
		h e^{i\theta}
	\end{pmatrix}e^{i k_y y}e^{i k_x x}
+
	\Xi\mathcal{R}_1^h\begin{pmatrix}
		i\\
		-h e^{-i\theta}
	\end{pmatrix}e^{i k_y y}e^{-i k_x x}\,,
	\label{wavefn_scatt_1_left}
\end{multline}
while in the region $x>0$,
\begin{equation}
	\psi_1^h(x>0,k,\theta) =\mathcal{T}\psi_{1,{\mathcal T}}= \Xi{\mathcal{T}_1^h}\begin{pmatrix}
		i\\
		h e^{i\theta}
	\end{pmatrix}e^{i k_y y}e^{i k_x x}\,,
	\label{wavefn_scatt_1_right}
\end{equation}
 {where $\Xi=(2A)^{-1/2}$ is the normalization factor with $A$ being the total surface area. We defined here $\bm k =(k_x,k_y) = (k\cos\theta,k \sin\theta)$ dropping the index $i$.}

{Note that even before computing the reflection and transmission coefficients it is clear why the potential $U_y\sigma_y$ does not affect the physics beyond an overall phase, as is seen from Eq.~\eqref{boundary_condition}. For the non-vanishing reflection coefficient, there must exist a non-vanishing matrix element of the scattering potential between the incoming and the reflected states. However, it is easy to verify that $\braket{\psi_i|\sigma_y|\psi_{\mathcal R}}=0$ for all incoming angles $\theta$. Consequently, this potential is reflectionless and does not lead to new phenomena.}
	
Utilizing the boundary condition in Eq.~\eqref{boundary_condition} for scattering states, we find the reflection and the transmission coefficients,
\begin{equation}
\mathcal{R}_1^h = -\frac{e^{i\theta} \sinh \eta ({h U_0} \sin \theta +{U_x})/v}{\eta\cos\theta   \cosh \eta+i  \sinh \eta ({h U_0}+{U_x} \sin \theta )/v},
\label{eq:R}
\end{equation}
and
\begin{equation}
\mathcal{T}_1^h = \frac{ \eta{\cos \theta }}{\eta{\cos\theta }  \cosh \eta+i \sinh \eta ({h U_0}+{U_x} \sin \theta )/v}.
\label{eq:T}
\end{equation}

{In the complementary process, for quasiparticles coming in from $x>0$, see Fig.~\ref{fig:scattering},}

\begin{equation}
\psi_2^h(x<0,k,\theta) ={\mathcal{T}}\psi_{2,{\mathcal T}}={\Xi}{\mathcal{T}_2^h}\begin{pmatrix}
i\\
h e^{i\theta}
\end{pmatrix}e^{i k_y' y}e^{i k_x' x},
\label{wavefn_scatt_2_left}
\end{equation}

and


\begin{multline}
\psi_2^h(x>0,k,\theta) = \equiv\psi_{2,i}+\mathcal{R}\psi_{2,{\mathcal R}}
\\
= {\Xi}\Bigg[\begin{pmatrix}
i\\
h e^{i\theta}
\end{pmatrix}e^{i k_x' x}
+
{\mathcal{R}_2^h}\begin{pmatrix}
i\\
-h e^{-i\theta}
\end{pmatrix}e^{-i k_x' x}\Bigg]e^{i k_y' y}\,.
\label{wavefn_scatt_2_right}
\end{multline}
{Once again, } imposing the boundary conditions, Eq.~\eqref{boundary_condition}, we obtain the reflection and transmission coefficients,


\begin{equation}
\mathcal{R}_2^h = -\frac{e^{i\theta} \sinh \eta ({h U_0} \sin \theta +{U_x})/v}{-\eta\cos\theta   \cosh\eta +i \sinh \eta ({h U_0}+{U_x} \sin \theta )/v},
\label{r2_U0Ux}
\end{equation}
and
%

\begin{equation}
\mathcal{T}_2^h = -\frac{ \eta{\cos \theta }}{-\eta{\cos\theta }  \cosh \eta+i \sinh \eta ({h U_0}+{U_x} \sin \theta )/v}.
\label{t2_U0Ux}
\end{equation}

{Inspection of Eqs.~\eqref{eq:R}-\eqref{eq:T} and Eqs.~\eqref{r2_U0Ux}-\eqref{t2_U0Ux} reveals several important observations. First, in the absence of magnetic scattering, $U_x=0$, quasiparticles exhibit Klein tunneling at normal incidence angles, i.e.  $|\mathcal{R}_{1,2}^h|^2=0$, as is expected for massless Dirac particles. When $U_x\neq 0$, there is a non-zero reflection probability at normal incidence.}

{Second, since $U_0$ appears only in combination with the helicity $h = \mbox{sgn}(E)$, the reflection and transmission coefficients are invariant under simultaneous transformation $E\rightarrow -E$ and $U_0\rightarrow -U_0$. This shows that quasiparticles above and below the Dirac point effectively feel opposite electrostatic potentials.}

{For all values of $U_0$ and $U_x$, the reflection probabilities, $|\mathcal{R}_1^h|^2=|\mathcal{R}_2^h|^2$ for each $\theta$.
Thus, we expect no current along the $x$-axis. To determine whether a current flows along the defect, we compare the reflection coefficients for the electrons incident at angles $\theta$ and $-\theta$. These are  not equal to each other whenever $U_x\neq 0$.  In a generic case both the magnitude and the phase of the reflection coefficients differ for these two angles. In the special situation of purely magnetic scattering ($U_0=0,U_x\neq 0$) the reflection probability is the same for $\theta$ and -$\theta$, but the phases of the coefficients $\mathcal{R}$ differ. Therefore, in all generality, in the presence of magnetic scattering, we expect that the time-reversal symmetry breaking is accompanied by charge currents along the defect line. In a strongly spin-momentum locked system such as the one we consider here, this also results in spin accumulation.  We discuss those in Secs.~\ref{sec:LDOS}-\ref{sec:charge_currents} below.}

\section{{Local density of states and Friedel oscillations} }
\label{sec:LDOS}

{Local density of states (LDOS) and its spin-resolved components are accessible, at least in principle, using scanning tunneling spectroscopies. We therefore compute their main features below. We start with the contribution of the bound states at a given energy, $\epsilon$. For each of the bound state branches, labeled by $\pm$, the $i$th spin component of LDOS is given by
\begin{equation}
  \rho_{i}^{b\pm}(\epsilon,x) =\frac{b}{2\pi} \int_{-\infty}^{\infty} dk_y \delta(\epsilon-v_\pm k_y) \braket{\psi^\pm|\sigma_i|\psi^\pm}\,,
\label{LDOS_bound_unsummed}
\end{equation}
where the wave functions are given in Sec.~\ref{sec:BS}, and $b$ is the system length used for normalizing the wave functions. The corresponding ``charge'' LDOS, $\rho^b(\epsilon,{x})$,  is obtained replacing the Pauli matrix, $\sigma_i$ by the identity matrix, and the total LDOS due to the bound states is the sum of the two contributions, $\rho^b_i(\epsilon,x) = \rho^{b+}_i(\epsilon,x)+ \rho^{b-}_i(\epsilon,x)$.}

\begin{figure}[t]
	\centering
	\includegraphics[width=0.48\textwidth]{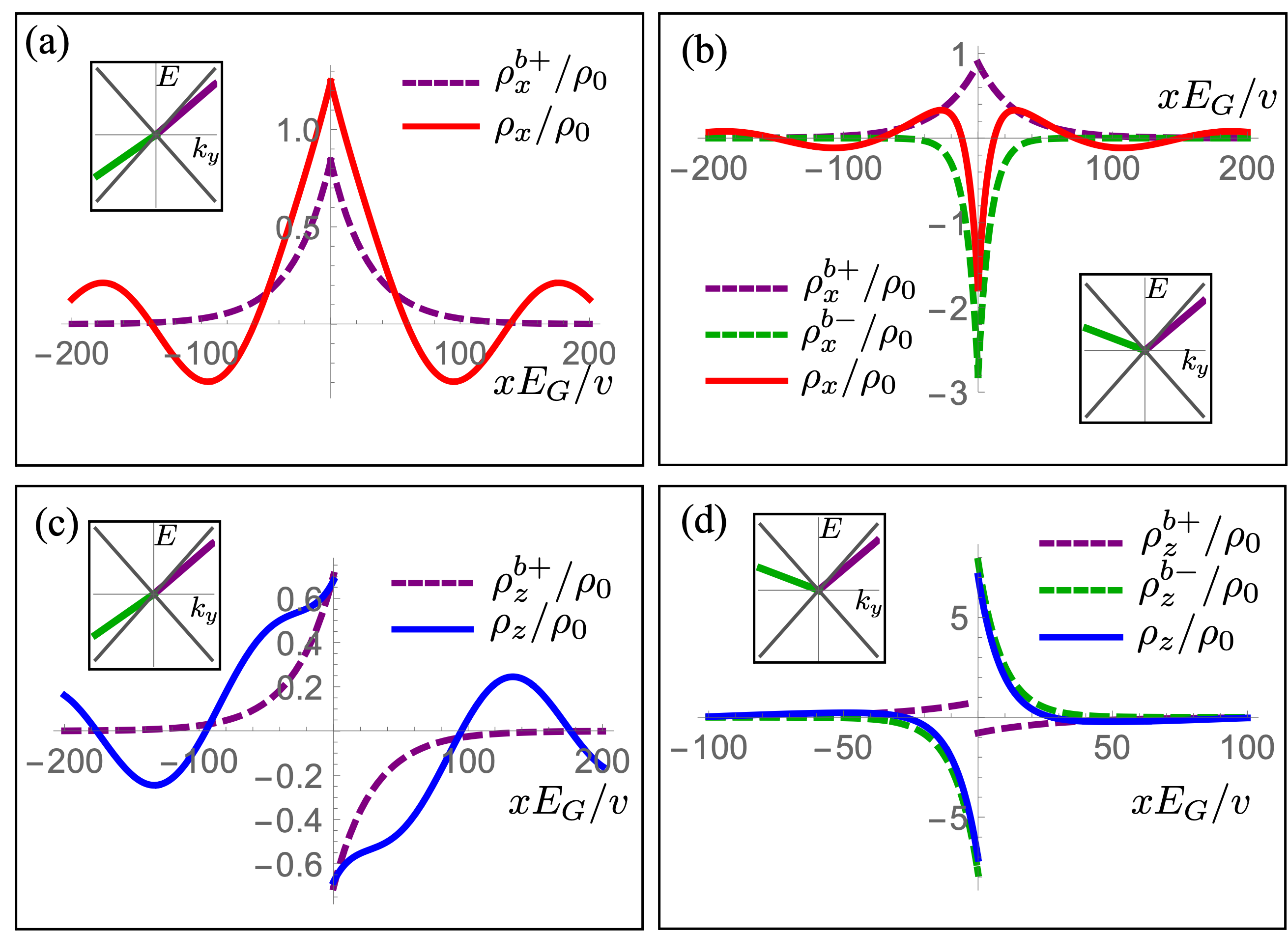}
	\caption{Spin resolved LDOS $\rho_{x}$ and $\rho_{z}$ for $\epsilon=0.02$. Panels (a), (c):  single bound state branch are for { $U_0 = -2$ and $U_x = 1.4$ {(we set $v=1$ as before)}. Panels (b),(d): two branches with opposite spin polarization are for $U_0 = -1$ and $U_x =0.3$. We show  the contribution of the bound states, $\rho^b_i$ and the total LDOS, $\rho_i$, including {the scattering contribution}. Insets of show the bound state dispersion for the corresponding cases.}}
	\label{spin_resolved_two_cases}
\end{figure}

{Upon momentum integration, $k_y=\epsilon/v_\pm$, and therefore we defined the energy-dependent inverse decay length, $\lambda_\pm^{(\epsilon)}=(\epsilon/v_\pm)\cos\alpha_\pm$ in analogy with Eq.~\eqref{eq:lambda}. If $\lambda_\pm^{(\epsilon)}<0$, no bound state exist at energy $\epsilon$ and hence there is no corresponding contribution to LDOS. While if $\lambda_\pm^{(\epsilon)}>0$, elementary integration yields}
%
\begin{subequations}
\begin{align}
\rho^{b\pm}(\epsilon,x) &= \frac{1}{2\pi}\frac{\lambda_{\pm}^{(\epsilon)}}{|v_\pm|} e^{-2\lambda_{\pm}^{(\epsilon)}|x|}\,,\\
\rho_{x}^{b\pm}(\epsilon,x) &= \frac{1}{2\pi} \frac{\lambda_\pm^{(\epsilon)}}{v} e^{-2\lambda_\pm^{(\epsilon)}|x|}  \text{sgn}(v_\pm) \label{rho_sx_U0Ux_bound}\,,\\
\rho_{z}^{b\pm}(\epsilon,x) &=-\frac{1}{2\pi} \frac{(\lambda_\pm^{(\epsilon)})^2}{\epsilon} e^{-2\lambda_\pm^{(\epsilon)}|x|} \text{sgn}\left(x{v_\pm}\right)\,. \label{rho_sz_U0Ux_bound}
\end{align}
\label{eqs:rho}
\end{subequations}
{Note that the spin-component normal to the plane changes sign across the $x=0$ line as expected from the symmetry arguments. The sign of the spin-projected LDOS depends on the dispersion of the bound states, $v_\pm$. In cases when two bound state branches exist at a given energy, their respective contributions may add ($v_-v_+>0$) or subtract ($v_-v_+<0$), and we give examples for both situations in Fig.~\ref{spin_resolved_two_cases}.}

{Scattering of the continuum states on the line defect produces Friedel oscillations in the LDOS. These oscillations are a consequence of the interference between the incoming and reflected waves in Fig.~\ref{fig:scattering} on the same side of the defect, and therefore are controlled by the reflection coefficients, ${\mathcal R}^h_{1,2}$, as shown in Appendix~\ref{sec:LDOS_appendix}. We evaluate them from the general expression,}
\begin{multline}
\rho^{s}_{j}(\epsilon,x) = \frac{A}{(2\pi)^2} \sum_{i = 1,2} \\ \int_{-\pi/2}^{\pi/2} d\theta \int_0^\infty dk k \braket{\sigma_j}_i  \delta(\epsilon- E(\bm{k})),
\label{LDOS_scatt_unsummed}
\end{multline}
{where $i = 1,2$ correspond to quasiparticles incident from $x\rightarrow -\infty$ and $x\rightarrow \infty$ respectively, see Fig.~\ref{fig:scattering}. At large distances, $k_\epsilon |x|\gg 1$, the integral can be evaluated analytically and has a familiar form, }
\begin{equation}
\frac{\Delta\rho^s(\epsilon,x)}{\rho_0}=F[U_0,U_x]\frac{\cos(2k_\epsilon |x|+\phi)}{(k_\epsilon |x|)^{3/2}},\,,
\end{equation}
{where we defined the deviation of the LDOS from the uniform value for an unperturbed Dirac cone, $\Delta\rho^s(\epsilon, x)=\rho^s(\epsilon, x)-\rho_0$, with $\rho_0=k_\epsilon/2\pi v$, and we introduced for convenience the momentum $k_\epsilon=|\epsilon|/v$. }Note that the $3/2$ power law for the total LDOS is different from the $1/2$ power characteristic of a two-dimensional electron gas, and agrees with Refs.~\cite{Crommie1993-wa,Biswas2011,An2012,Liu2012}. The corresponding spin-resolved LDOS,
\begin{equation}
\frac{\rho^s_{x,z}(\epsilon,x)}{\rho_0}=F_{x,z}[U_0,U_x]\frac{\cos(2k_\epsilon |x|+\phi_{x,z})}{(k_\epsilon |x|)^{1/2}},
\end{equation}
vanishes unless $U_x\neq 0$. Functions  $F,F_x,F_z$ and the phases $\phi,\phi_x$, $\phi_z$ are given in Eqs.~\eqref{LDOS_scatt},~\eqref{xLDOS_scatt4} and~\eqref{zLDOS_scatt4}{ of the appendix. Their general form is not crucial for our analysis.}

{Note that the spin resolved LDOS decays slower than the net LDOS. Recall that the Friedel oscillations arise from the interference between incident and reflected waves, and the asymptotic form at $k_\epsilon |x|\gg 1$ is dominated by near backscattering. Spin momentum locking in TIs ensures that as $\theta\rightarrow 0$, the overlap $\braket{\psi_{\mathcal{R}}|\psi_{i}}\rightarrow 0$, reducing the interference effects and leading to a faster decay of $\rho^s$.  At the same time $\braket{\psi_{\mathcal{R}}|\sigma_{x,z}|\psi_{i}}$ does not vanish in the same limit, ``protecting'' the 1/2 power law  for $\rho^s_{x,z}$. }

{In the limit $U_x\gg U_0$ and $U_x \gg 1$, $|\mathcal{R}^h_{1,2}|\rightarrow 1$ 
the integrals can be evaluated exactly at arbitrary values of $x$ to give}
\begin{subequations}
\begin{align}
\rho^{s}(\epsilon,x) &= - \frac{|\epsilon|}{2\pi v^2}  \frac{J_1(2k_\epsilon x)}{2k_\epsilon x},\\
\rho^{s}_{x}(\epsilon,x) &= \frac{\text{sgn}(U_x)}{2\pi} \frac{\epsilon}{v^2}  J_1(2k_\epsilon |x|), \label{LDOS_x_large_Ux}\\
\rho^{s}_{z}(\epsilon,x) &= \frac{\text{sgn}(x U_x)}{2\pi} \frac{k_\epsilon}{v}  \left(\frac{J_1(2k_\epsilon x)}{2k_\epsilon x}- J_2(2k_\epsilon x) \right),
\end{align}
\end{subequations}
where $J_1$ and $J_2$ are Bessel functions of first kind.

Since $\lambda_\pm/ k_\epsilon\sim 1$ for most values of $U_0$ and $U_x$, at distances larger than $1/k_\epsilon$ the Friedel oscillations determine the LDOS.
Close to the line we evaluate the LDOS numerically, and find that the bound states often, but not always dominate,{ with the details depending} on the specific values of $U_0$ and $U_x$, see appendix \ref{sec:bound_scatt_comparison}. This holds at all energies since $\lambda_\pm$ scales linearly with $\epsilon$.

{Characteristic behavior of LDOS is shown in Fig.~\ref{spin_resolved_two_cases}. As discussed above, near the defect line the bound state LDOS depends on whether we have one or two branches at a given energy, and we show the corresponding cases from Fig.~\ref{dispersion} as insets for reference. If only a single branch exists, Fig.~\ref{dispersion}(b),  LDOS shows  a clear exponential decay superimposed on Friedel oscillations,  Fig.~\ref{spin_resolved_two_cases}(a).  If there are two branches, the LDOS values for the spin component $s_x \propto v_\pm$  add (subtract) when $v_+v_->0$ ($v_+v_-<0$), with examples in Fig.~\ref{dispersion}(d)(Fig.~\ref{dispersion}(a)).
For $v_+v_-<0$, since $\lambda_+\neq\lambda_-$,  the more localized state dominates near $x=0$ and its counterpart yields the opposite spin orientation at intermediate distances, see Fig.~\ref{spin_resolved_two_cases}(b). The situation for $\rho_z$ is similar, but must incorporate the sign change at $x=0$, see Fig.~\ref{spin_resolved_two_cases}(c,d). At least in principle these contributions to  LDOS can be observed using spin-polarized STM.}



 \section{\label{sec:charge_currents}Spin Accumulation and Charge Currents.}

These LDOS features lead to spatially varying spin accumulation. The spin density (per unit area) at $T=0$ is given by 
\begin{equation}
	{	\mathcal{S}_{x,z} (x) = \int_{-\Lambda}^{\mu} \left[\rho^s_{x,z}(\epsilon,x)+\rho^b_{x,z}(\epsilon,x)\right]d\epsilon\equiv \mathcal{S}_{x,z}^s+\mathcal{S}_{x,z}^b\,,}
	\label{eq:spin_acc}
\end{equation}
where $\mu$ is the chemical potential and $\Lambda$ is the high energy cutoff that we take to coincide with the top of the valence band, $\Lambda = E_G/2$.
{Complete details of the calculation are given in Appendix~\ref{app:spin_acc}, and here we only emphasize the important features of the results. The net contribution of the scattering states arises from the difference between the spin accumulation due to states below and above the Dirac point,
\begin{equation}
	\mathcal{S}_{x,z}^s =	\mathcal{S}_{x,z}^{s,\mu}	-\mathcal{S}_{x,z}^{s,-\Lambda}\,,
\end{equation}
where each of the terms has the form familiar from the Friedel oscillations ($\beta=\mu,-\Lambda$) 
\begin{equation}
		\mathcal{S}_{x,z}^{s,\beta} = \left(\frac{\beta}{v}\right)^2K_{x,z}^\beta[U_0,U_x] \frac{\cos(2|\beta x|/v + \tilde\phi_{x,z})}{(|\beta x|/v)^{3/2}}\,,
		\label{eq:spin_acc_scatt}
\end{equation}
with $K_{x,z}^\beta$ and $\tilde\phi_{x,z}$ given in Appendix~\ref{app:spin_acc}.  Note that the long-range decay of each contribution goes as $\mathcal{S}_{x,z}^{s,\beta} \propto |x|^{-3/2} l_\beta^{-1/2}$, where $l_\beta=1/k_\beta=v/|\beta|$ is the characteristic length scale for the high energy cutoff and the chemical potential, respectively.}

{Similarly, the cutoff in the integration of the spin accumulation due to the bound states, $\beta$, depends on whether the corresponding branch is below ($\beta=-\Lambda$) or above ($\beta=\mu$) the Dirac point.
Naively, it would seem that the contribution of the bound states is much more localized. However, since the localization length diverges as $\epsilon\rightarrow 0$, namely  $\lambda^{(\epsilon)}_\pm\propto \epsilon/v$, the low energy bound states provide a long-range tail to the accumulated spin density. Integration in Eq.~\eqref{eq:spin_acc} with the densities from Eq.~\eqref{eqs:rho} gives}
\begin{equation}
\mathcal{S}_{x,z}^b\sim \frac{(\beta/v_\pm)^2}{|\lambda_\pm^{(\beta)} x|^2},
\end{equation}
where $\lambda_\pm^{(\beta)}$ is evaluated at $\beta = \mu,-\Lambda$ depending on whether  the branch is above or below the Dirac point, see appendix \ref{app:spin_acc} for full expressions. {For a generic case when the bound state is not close to merging with the scattering continuum, $v_\pm\lesssim v, \lambda_\pm^{(\beta)}\sim \beta/v$, the contribution of the bound states simply decays as $|x|^{-2}$. While it is notable that the decay of the spin accumulation due to the bound states is non-exponential, in the regime of the validity of Eq.~\eqref{eq:spin_acc_scatt} ($|x|/l_\beta\gg 1$), the scattering states still dominate as $\mathcal{S}_{x,z}^s/ \mathcal{S}_{x,z}^b\sim (|x|/l_\beta)^{1/2}$, albeit not as strongly as one would naively expect.}
{The out-of-plane spin density, ${\mathcal S}_z$, creates a magnetic field, and may be detected in magnetometry measurements such as SQuID. The in-plane magnetization may potentially be detected optically, from the magneto-optical measurements.}


\begin{figure}[t]
\centering
\includegraphics[width=0.45\textwidth]{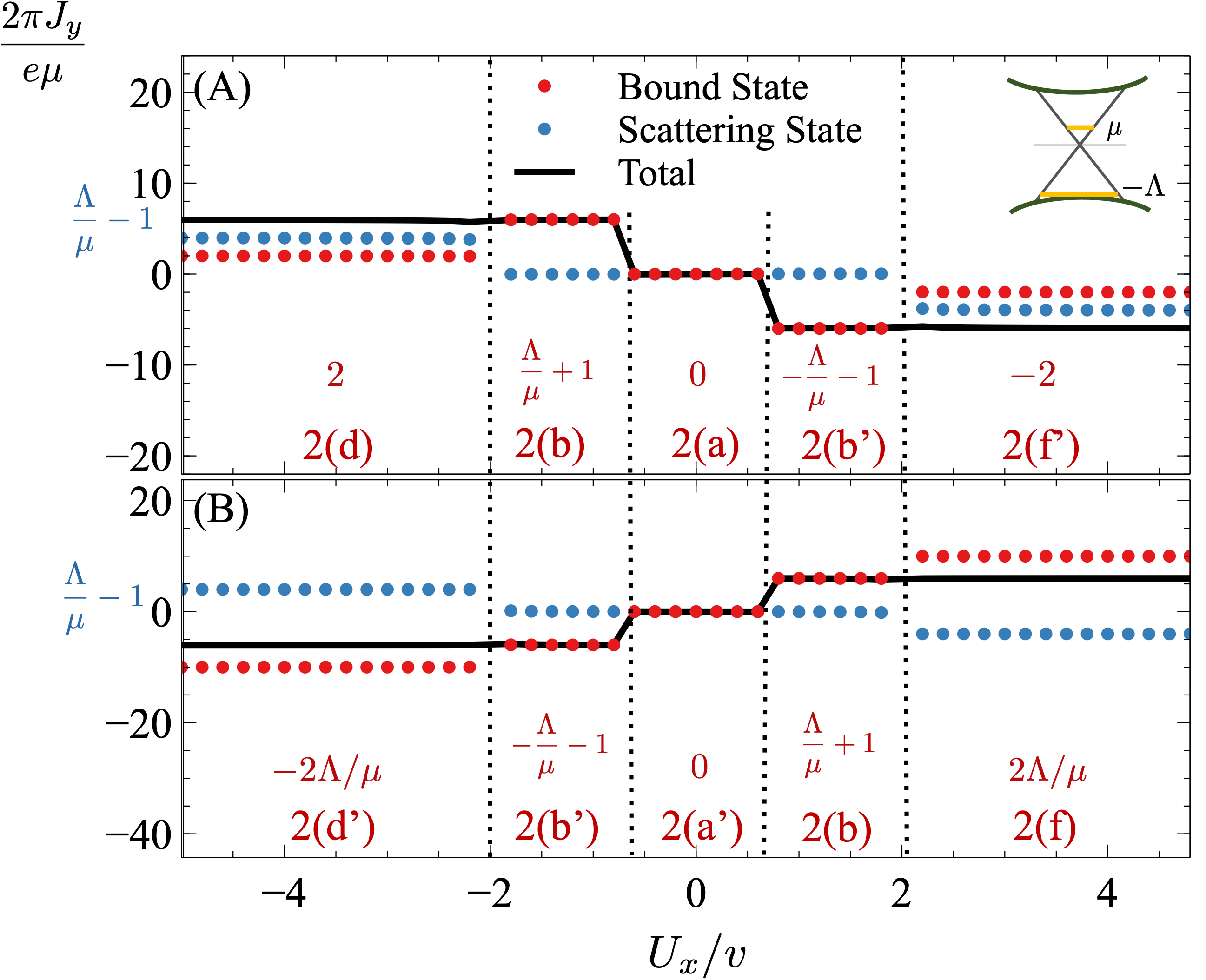}
\caption{Charge current along the defect showing stepwise dependence on $U_x$. We take $\mu =0.1$, $v=1$, $\Lambda = 0.5$. Panel (A): $U_0 = 2$; panel (B): $U_0 = -2$. Panel (A) inset: sketch of the Dirac cone and bulk bands with $\Lambda$ and $\mu$ identified.  We note the combinations of $\Lambda$ and $\mu$ that enter the net current (see text) and refer to the corresponding dispersion in Fig.~\ref{dispersion}. }
\label{Charge_current_vary_Ux}
\end{figure}

 Since the current operator for Dirac systems is proportional to the spin, e.g. $j_y=e\delta H/\delta k_y =ev \sigma_x$, spin accumulation leads to net charge currents. Note $\mathcal{J}_x \sim \langle \sigma_y\rangle=0$ while the current $\mathcal{J}_y(x)$ flows along $\mbox{sgn}(U_x)\bm{\widehat x}\times\bm{\widehat z}$, parallel to the defect, and has a spatial profile  similar to that of ${\mathcal S}_x(x)$ above. 

The net current at $T=0$ is obtained by summing over the occupied states,
\begin{equation}
{J}_y = ev \int_{-\Lambda}^{\mu}d\epsilon\int_{-\infty}^{\infty} dx   \rho_x(\epsilon, x)
\end{equation}

{Using Eq.~\eqref{rho_sx_U0Ux_bound}, and performing the spatial integration results in energy-independent integrand, $\text{sgn}(v_\pm)/2\pi v$, for the energy integral.} Thus, for $\mu>0$, the bound state currents are independent of the values of the velocity, and are $\text{sgn}(v_\pm)e\mu /2\pi$ and $\text{sgn}(v_\pm)e\Lambda/2\pi$ for branches above and below Dirac point, respectively. For $\mu<0$, each branch below the Dirac point contributes $\text{sgn}(v_\pm)e(\mu+\Lambda)/2\pi$.
Therefore, the bound state contribution to $J_y$ changes stepwise with the magnetic potential $U_x$ as the branches evolve according to Fig.~\ref{dispersion}.

The contribution of the scattering states is small for $|U_x|<|U_0|$ since the scattering anisotropy in $\mathcal{R}$, Eq.~\eqref{eq:R}, is weak, and the currents due to electrons incoming at the angles $\theta$ and $-\theta$ nearly compensate.  In the opposite limit,  $|U_x|\gg|U_0|$ the reflection coefficient $|{\mathcal R}|\rightarrow 1$ irrespective of $\theta$.
Performing spatial integration over $\rho^s_x$ (Eq.~\eqref{rho_sx_U0Ux_bound}) and summing over filled states, we find { that the limiting value of the current due to the scattering state is also insensitive to the magnitude of the magnetic scattering potential, namely
\begin{equation}
	J_{y,0}^s= \text{sgn}(U_x)\frac{e (|\mu|-\Lambda)}{2\pi}\,.
\end{equation}
}
This saturated ($|U_x|$-independent) value is evident already at moderate values of $U_x$ in Fig.~\ref{Charge_current_vary_Ux}, where we evaluated all currents numerically, and found that the total current also varies nearly stepwise with $U_x$.
{One of the  values of the $U_x$ where the current changes nearly discontinuously, marked by vertical dashed lines, corresponds  to $|U_x|=|U_0|$ (2 in our case). Here $\eta=\sqrt{U_x^2-U_0^2}/v$ changes from real to imaginary and therefore the amplitude of the Friedel oscillations, as well as spin accumulation change abruptly, see Appendix~\ref{app:spin_acc}. The other discontinuity in the current due to the scattering states  occurs when  sign of bound state velocity of one of the branches changes. For $|U_x|<|U_0|$ we find this to happen numerically at $U_x^\star/v\approx\pm 0.66$ for values in Fig.~\ref{Charge_current_vary_Ux}.  Setting $v_\pm=0$ in  Eq.~\eqref{v_eff} with $|U_x|<|U_0|$, this corresponds to $U_x\mp U_0\cos(\sqrt{U_0^2-U_x^2}/v)=0$. Expansion to the second order in $U_x/U_0$ yields a close approximate solution,
for the critical value of the scattering potential when the dispersion of the bound states becomes flat,
\begin{equation}
\frac{U_x^\star}{v} \approx \frac{\pm(-1+\sqrt{1-(U_0/v)\sin (2U_0/v)})}{\sin (U_0/v)} \approx 0.64.
\end{equation}

Thus, both the current due to the scattering states and that due to the bound states vary discontinuously with the value of the magnetic scattering potential, $U_x$, and hence the total current varies stepwise with $U_x$. In the next section we discuss how this current can be controlled. It is important to note, however, that these currents are dissipationless, and therefore are difficult to detect in transport measurements. Instead, they can be identified by the magnetic fields they generate or via optical measurements.
}

\section{\label{sec:control}Control of spin-accumulation and charge currents} 
{One of our main findings is that magnetic scattering from the spins aligned with the defect line ($U_y$) and those normal to it in the plane ($U_x$) has very different consequences for the observable spin textures and currents. This opens an avenue for on-demand control of the bound state dispersion and scattering properties of extended  states. If the magnetic scattering is due to the classical spins ($S$) on the defect, the direction of the magnetic moments in the plane determines the ratio $U_x/U_y \propto S_x/S_y$. In the absence of in-plane magnetic anisotropy a magnetic field, $\bm B$, applied along the surface controls the direction of spins along the defect line, $ B_x/B_y \approx S_x/S_y$. Assuming that the potential scattering is insensitive to the applied field, and recalling that $U_y$ is irrelevant for physical observables, we are led to conclude that rotating the field with respect to the line defect effectively changes the ratio $U_x/U_0$. This, in turn, controls the spin accumulation and charge currents.}

{Now we show that such a field does not alter the electronic properties or the boundary conditions that we used to reach our conclusions. 
Orbital coupling shifts the momentum of an electron, ${\bm k} \rightarrow {\bm k} - e{\bm A}$. For the field along the surface, choosing the vector potential along the $z$-axis, ${\bm A} = (0,0,B_x y - B_y x)$, leaves $k_x$ and $k_y$ unchanged. For the same in-plane field we must also account for the Zeeman term $-\frac{g \mu_B}{\hbar} {\bm \sigma} \cdot {\bm B}$, where  $\mu_B$ is the Bohr magneton, and $g$ is the gyromagnetic factor.
Zeeman contribution  results in shift in both $\hat{k_y}$ and $\hat{k_x}$ ~\cite{Kane_RMP_2010,Zhang_RMP_2011}, $\hat{k_x} \rightarrow \hat{k_x} + \frac{g \mu_B}{2} B_y $ and $\hat{k_y} \rightarrow \hat{k_y} - \frac{g \mu_B}{2} B_x $. This feature is a consequence of the linearity of the Hamiltonian in the momentum $\bm k$, see Eq.~\eqref{full_hamiltonian}. 
The shift in $\hat{k_x}$  appears as an overall phase, so can also be gauged away by making the choice ${\bm A} = (\frac{g \mu_B}{2e} B_y,0,B_x y - B_y x)$. In turn, the shift of the momentum  $\hat{k}_y$ simply relocates the entire spectrum, including the Dirac point, to a finite momentum. Crucially, because of the same linearity of the Hamiltonian in ${\bm k}$, this shift does not affect the current operator. Since the spin structure of the states is also insensitive to the location of the Dirac point in the momentum space, none of the physical observables depend on the in-plane magnetic field. }

Also note that boundary condition remains unchanged, since the integral across the defect that was performed to arrive to Eq.~\eqref{boundary_condition} depends only on the terms that are singular at the defect, which orbital coupling and Zeeman term are not. {We therefore conclude that, if an applied magnetic field is rotated in the surface plane, it tunes the value of $U_x$ and, consequently, the ratio $U_x/U_0$. The maximal ratio is achieved for the field normal to the defect line, while the field along the defect line removes the observable effects of magnetic component of the scattering.  Hence the  discreteness of the currents as a function of $U_x$, shown in Fig.~\ref{Charge_current_vary_Ux}, directly translates into discrete jumps as a function of the field direction. }

\section{Disorder broadening.}
Since extended and bound states coexist at different $k_y$ for the same energy $E$, randomly located point impurities mix the two. {We estimate the broadening of the bound states in the Born approximation. To the second order in the scattering potential,
	\begin{equation}
		\Gamma= 2\pi \sum_{\vec{k}} |\braket{\psi_{s}|\hat{V}|\psi_b}|^2\delta(\epsilon -E(\bm{k})),
		\label{broadening_def}
	\end{equation}
	where $\ket{\psi_b}$ and $\ket{\psi_s}$ are the bound state and scattering state wavefunctions respectively and $\hat{V}$ is the impurity potential, which we take to be a superposition of randomly distributed point-like scatterers of strength $V_0$. We find (see Appendix~\ref{sec:born_app} for  details)
	\begin{equation}
		\Gamma = \frac{n_{imp} V_0^2 |\epsilon|}{v^2}F(v_\pm/v).
	\end{equation}
Here $n_{imp}$ is the impurity concentration. The result is intuitively clear as the broadening  is proportional to density of extended states available for scattering, $\rho_0 = |\epsilon|/2\pi v^2$, and the usual Born factor  $n_{imp}V_0^2$. The spin-momentum locking and other details of the states are captured solely by the appearance of a monotonic but bounded function, $F(v_\pm/v) \leq \pi$, which depends on the mismatch between $v_\pm$ and $v$, see Eq.~\eqref{F_eq}. This result means that the bound states are parametrically well defined at sufficiently clean surfaces, under the condition $n_{imp} V_0^2/v^2\ll 1$.}

\section{Discussion and conclusions.}
We showed that linear defects with magnetic component of scattering at surfaces of topological insulators support spin-polarized  bound states, whose signatures are accessible by local scanning probes. {Spin structure of the bound states combines with the asymmetric scattering of the extended quasiparticles, due to breaking of time-reversal symmetry, and results in macroscopic spin accumulation and the flow of non-dissipative charge currents along the defect. Our most important conclusions are that the magnetic moments of the scattering centers along and normal to the defect line play very different roles, and therefore varying the angle between an external magnetic field applied along the surface and the defect line effectively controls the strength of magnetic scattering. Since the charge currents vary stepwise as a function of the strength of magnetic scattering, the same stepwise dependence will appear as a function of the  field direction in the plane.}

Above we used a continuum long-wavelength Hamiltonian.  In lattice models the bound states merge with either valence or conduction band~\cite{Xu2018} at momenta comparable to the size of the Brillouin zone. The resulting non-linearity of the dispersion will modify the values of the net magnetic moment and the current, and wash out the sharp transitions between the plateaus in Fig.~\ref{Charge_current_vary_Ux}, but our main conclusions remain unaffected. Similarly, weak hexagonal modulation of the Dirac cone changes the quantitative details but not the qualitative behavior found here.

{It is important to note that the currents we find are a feature of the ground state, and therefore dissipationless (flow in the absence of external bias). Consequently, they cannot be easily measured using standard transport techniques and geometries. Instead, these currents will be most easily accessible and detectable via the magnetic fields they produce. In this context, our work motivates studies of patterned networks of line defects, where the desired spatial distribution of these currents and the associated fields can be created. In this context our work is a part of a bigger effort of defect engineering of surface and interface properties. }




\begin{acknowledgments}
This work started at KITP Santa Barbara where it was supported in part by the National Science Foundation under Grant No. NSF PHY-1748958. and was supported by NSF via Grant No.
DMR-141074. We are grateful to D.~E.~Sheehy, W.~A.~Shelton, and J.~H.~Wilson for discussions.
\end{acknowledgments}

\appendix

\section{{Wavefunction of the bound states.}}


We look for eigenstates of the Hamiltonian, Eq.~\eqref{full_hamiltonian} of the main text, of the form

\begin{equation}
\psi(x,k_y) =
\begin{pmatrix}
C\\
D
\end{pmatrix}e^{ik_y y} e^{-\lambda|x|} \sqrt{\frac{\lambda}{b}},
\end{equation}
where $b$ is normalization length along $y$, and $\lambda >0$ for solutions to be normalizable along $x$. By substituting it in the hamiltonian we find,

\begin{equation}
v\begin{pmatrix}
0 & k_y-\lambda\text{sgn}(x)\\
k_y+\lambda\text{sgn}(x) & 0
\end{pmatrix}\begin{pmatrix}
C\\
D
\end{pmatrix}
= E\begin{pmatrix}
C\\
D
\end{pmatrix}.
\end{equation}

The eigenvalue for the above equation is $E=\pm v\sqrt{k_y^2-\lambda^2}$.

{For real eigenvalues $E$, $\lambda<|k_y|$. We therefore  introduce parameters $\alpha_\pm$ such that energy eigenvalue $E_\pm = \pm vk_y\sin\alpha_\pm$ and $\lambda_\pm = k_y \cos\alpha_\pm$. Then solving for the eigenfunctions we obtain the wavefunctions in Eqs.~\eqref{wavefn_bound_right} and~\eqref{wavefn_bound_left}. Imposing the boundary condition (Eq.~\eqref{boundary_condition} of the main text) we  obtain the energy eigenvalues,
\begin{equation}
E =- v k_y \left(\frac{U_x \mp U_0\cosh\eta}{U_0\mp U_x \cosh\eta}\right),
\end{equation}
and Eqs.~\eqref{eq:alpha} and~\eqref{eq:lambda} follow.}

\section{\label{sec:LDOS_appendix}Local Density of States: Scattering States}

Scattering state LDOS is given by Eq.~\eqref{LDOS_scatt_unsummed} with $\sigma_i$ replaced by identity matrix.

\begin{multline}
\rho^{s}(\epsilon,x) = \frac{A |\epsilon|}{(2\pi)^2v^2} \sum_{i = 1,2}  \int_{\pi/2}^{\pi/2} d\theta |\psi_i^h (k_\epsilon,\theta)|^2\,,
\label{LDOS_def}
\end{multline}
where $A$ is the area of the surface, and appears for normalization. $k_\epsilon = |\epsilon|/v$ and we have suppressed spatial dependence of wavefunction for brevity. Sum over $1$ and $2$ corresponds to particle coming in from left and right respectively. Since the two are not coherent, we sum the corresponding amplitudes. For $x<0$, 

%
%


%
\begin{multline}
\int_{-\pi/2}^{\pi/2} d\theta (|\psi_1^h(k_\epsilon,\theta)|^2 + |\psi_2^h(k_\epsilon,\theta)|^2) = \\ \frac{1}{A}\left(2\pi + 2\mathrm{Re} \left(\int_{-\pi/2}^{\pi/2} d\theta \mathcal{R}_1^h e^{-2ik_\epsilon x\cos\theta} i e^{-i\theta} \sin\theta \right)\right).
\label{wvfn_sum}
\end{multline}

{It is worth emphasizing that it is the reflection coefficient, ${\mathcal R}$ that determines the interference between the incoming and the outgoing states that leads to Friedel oscillations. }In the limit $k_\epsilon |x|>>1$, the dominant contribution comes only from near the stationary points within the integration interval. However, $\mathcal{R}_1^h i e^{-i\theta} \sin\theta$ vanishes at $\theta =0$. Instead we use a generalized version of stationary phase approximation~\cite{Bhattacharya1979} and the leading order contribution is computed to be, 



\begin{equation}
\frac{\Delta\rho^s(\epsilon,\bm r)}{\rho_0}=F[U_0,U_x]\frac{\cos(2k_\epsilon |x|+\phi)}{(k_\epsilon |x|)^{3/2}}
\end{equation}
where $\Delta \rho^{s}(\epsilon,x) = \rho^s(\epsilon,x)-\rho_0$, $\rho_0 =  k_\epsilon/2\pi v$,

\begin{align}
F[U_0,U_x] = \frac{\sqrt{C_1^2+S_1^2}}{2\sqrt{\pi}}, {\hspace{5mm}}& \phi = -\frac{3\pi}{4} - \tan^{-1}\left(\frac{S_1}{C_1}\right),
\label{LDOS_scatt}
\end{align}

\begin{equation}
C_1=-\frac{\eta^2\sinh^2\eta(U_x^2  \cosh^2\eta + U_0^2)}{(\eta^2  \cosh^2\eta + U_0^2\sinh^2\eta)^2},
\end{equation}
and

\begin{equation}
S_1=-\text{sgn}(\epsilon)\frac{U_0\eta \sinh\eta \cosh\eta (\eta^2 - U_x^2 \sinh^2\eta)}{(\eta^2  \cosh^2\eta + U_0^2\sinh^2\eta)^2}.
\end{equation}




%
%

%


\section{\label{sec:spin_LDOS_appendix}Spin-resolved Local Density of States}

{

\subsection{Continuity of $\rho_x$ and Discontinuity of $\rho_z$}

Note that despite the rotation and spinor magnitude change at the defect, see Eq.~\eqref{boundary_condition}, $\rho_x$ is continuous across the defect and $\rho_z$ is flips direction at the defect. To demonstrate  this, we explicitly calculate $\braket{\psi^{\pm}(x=0^-,k_y)|\sigma_{x,z}|\psi^{\pm}(x=0^-,k_y)}$ and $\braket{\psi^{\pm}(x=0^-,k_y)|\mathcal{M}^\dagger \sigma_{x,z}\mathcal{M}|\psi^{\pm}(x=0^-,k_y)}$ for $U = U_0\delta(x)$ case. Setting $U_x=0$, the wavefunction

\begin{equation}
\ket{\psi^{\pm}(x=0^-,k_y)} = \sqrt{\frac{1\mp \sin U_0}{2}}\begin{pmatrix}
1\\
\pm \frac{\cos U_0}{1\mp \sin U_0}
\end{pmatrix}
\sqrt{\frac{\lambda_\pm^{(\epsilon)}}{b}} e^{i k_y y}.
\end{equation}

Then
\begin{equation}
\braket{\psi^{\pm}(x=0^-,k_y)|\sigma_x|\psi^{\pm}(x=0^-,k_y)} = \pm \cos U_0\lambda_\pm^{(\epsilon)}/b,
\end{equation}

and

\begin{equation}
\braket{\psi^{\pm}(x=0^-,k_y)|\sigma_z|\psi^{\pm}(x=0^-,k_y)} = \mp \sin U_0\lambda_\pm^{(\epsilon)}/b.
\end{equation}

Now we find

\begin{multline}
\mathcal{M}\ket{\psi^{\pm}(x=0^-,k_y)} =
\begin{pmatrix}
\cos U_0 & \sin U_0\\
-\sin U_0 & \cos U_0
\end{pmatrix}
\begin{pmatrix}
1\\
\pm \frac{\cos U_0}{1\mp \sin U_0}
\end{pmatrix}
\\ \sqrt{\frac{\lambda_\pm^{(\epsilon)}}{b}} e^{i k_y y} \sqrt{\frac{1\mp \sin U_0}{2}}
\end{multline}

\begin{equation}
= \begin{pmatrix}
\frac{\cos U_0}{1\mp \sin U_0}\\
\pm 1
\end{pmatrix} \sqrt{\frac{\lambda_\pm^{(\epsilon)}}{b}} e^{i k_y y} \sqrt{\frac{1\mp \sin U_0}{2}}
\end{equation}

Explicit evaluation of the expectation values shows that
\begin{multline}
\braket{\psi^{\pm}(x=0^-,k_y)|\mathcal{M}^\dagger \sigma_x\mathcal{M}|\psi^{\pm}(x=0^-,k_y)} \\= \pm \cos U_0 \lambda_\pm^{(\epsilon)}/b,
\end{multline}

and

\begin{multline}
\braket{\psi^{\pm}(x=0^-,k_y)|\mathcal{M}^\dagger \sigma_z\mathcal{M}|\psi^{\pm}(x=0^-,k_y)} \\= \pm \sin U_0 \lambda_\pm^{(\epsilon)}/b.
\end{multline}

Thus, we see that $\braket{\sigma_x}$ has remained the same across the defect while $\braket{\sigma_z}$ has flipped sign. This is a consequence of mirror symmetry about $y$-axis of the Hamiltonian which the bound eigenstates have inherited.
}

\subsection{Scattering States LDOS}
Spin-resolved LDOS is defined  in general in Eq.~\eqref{LDOS_scatt_unsummed} of the main text, and here we focus separately on the $x$ and $z$ components.

%
%

%

\subsubsection{Spin-resolved LDOS, $\rho^{s}_{x}(\epsilon,x)$}

%
%
%
%
%
%
%
%
%
%
%
%
%

We first consider $\rho^{s}_{x}$, which  is given by,
\begin{multline}
\rho^{s}_{x}(\epsilon,x<0) \\= \frac{2}{(2\pi)^2} \frac{|\epsilon|}{v^2} \int_{-\pi/2}^{\pi/2}  \text{Re} ( \mathcal{R}_1^h h i e^{-i\theta} e^{-2i k_\epsilon x \cos\theta}) d\theta.
\label{LDOS_x_ref_prob_dep}
\end{multline}

Recall that $\rho_x^s$ is symmetric about $x=0$ due to mirror symmetry about $y$-axis. In the limit $k_\epsilon|x|>>1$, the dominant contribution is from angles near $\theta=0$. Using stationary phase approximation~\cite{Bhattacharya1979},

%

\begin{equation}
\frac{\rho^s_x(\epsilon,\bm r)}{\rho_0}=F_x[U_0,U_x]\frac{\cos(2k_\epsilon |x|+\phi_x)}{(k_\epsilon |x|)^{1/2}},
\label{xLDOS_scatt1}
\end{equation}
where

\begin{align}
F_x[U_0,U_x] = \text{sgn}(\epsilon)\frac{\sqrt{C_2^2+S_2^2}}{\sqrt{\pi}}, {\hspace{0mm}}& \phi_x = -\frac{\pi}{4} - \tan^{-1}\left(\frac{S_2}{C_2}\right),
\label{xLDOS_scatt4}
\end{align}

\begin{equation}
C_2 = -\text{sgn}(\epsilon)\frac{U_x U_0 \sinh^2\eta}{\eta^2\cosh^2\eta + U_0^2 \sinh^2\eta},
\label{xLDOS_scatt2}
\end{equation}
and

\begin{equation}
S_2 = \frac{U_x \eta \sinh\eta \cosh\eta}{\eta^2\cosh^2\eta + U_0^2 \sinh^2\eta}.
\label{xLDOS_scatt3}
\end{equation}

\vspace{10mm}

\subsubsection{Spin-resolved LDOS, $\rho^{s}_{z}(\epsilon,x)$}

The $z$ components of the spin-LDOS $x<0$ is given by

\begin{multline}
\rho^{s}_{z}(\epsilon,x<0)  \\= \frac{1}{(2\pi)^2} \frac{|\epsilon|}{v^2} \int_{-\pi/2}^{\pi/2} 2 \text{Re} (\cos\theta \mathcal{R}_i^h e^{-i\theta} e^{-2i \epsilon x \cos\theta/v}) d\theta.
\label{LDOS_z_ref_prob_dep}
\end{multline}

Recall that $\rho_z^s$ is anti-symmetric about $x=0$ due to mirror symmetry about the $y$-axis. In the limit $k_\epsilon|x|>>1$ the decay law for $\rho_z^s$ is similar to $\rho_x^s$,

%

\begin{equation}
\frac{\rho^s_z(\epsilon,\bm r)}{\rho_0} = F_z[U_0,U_x]\frac{\cos(2k_\epsilon |x|+\phi_z)}{(k_\epsilon |x|)^{1/2}}
\label{zLDOS_scatt1}
\end{equation}
where

\begin{multline}
F_z[U_0,U_x] = \frac{\text{sgn}(x)\sqrt{C_3^2+S_3^2}}{\sqrt{\pi}} , {\hspace{3mm}}\\ \phi_x = -\frac{\pi}{4} - \tan^{-1}\left(\frac{S_3}{C_3}\right),
\label{zLDOS_scatt4}
\end{multline}

\begin{equation}
C_3 = -\frac{U_x \eta \sinh\eta \cosh\eta}{\eta^2\cosh^2\eta + U_0^2 \sinh^2\eta},
\label{zLDOS_scatt2}
\end{equation}
and

\begin{equation}
S_3 = -\text{sgn}(\epsilon)\frac{U_x U_0 \sinh^2\eta}{\eta^2\cosh^2\eta + U_0^2 \sinh^2\eta}.
\label{zLDOS_scatt3}
\end{equation}


{In the main text, we stated that the Friedel oscillations in spin-resolved LDOS decays slower that LDOS. Eqs.~\eqref{xLDOS_scatt1} and \eqref{zLDOS_scatt1} explicitly show this behavior.}

\section{\label{sec:bound_scatt_comparison}Comparison of Bound and Scattering States}


\begin{figure}[H]
\centering
\includegraphics[width=0.4\textwidth]{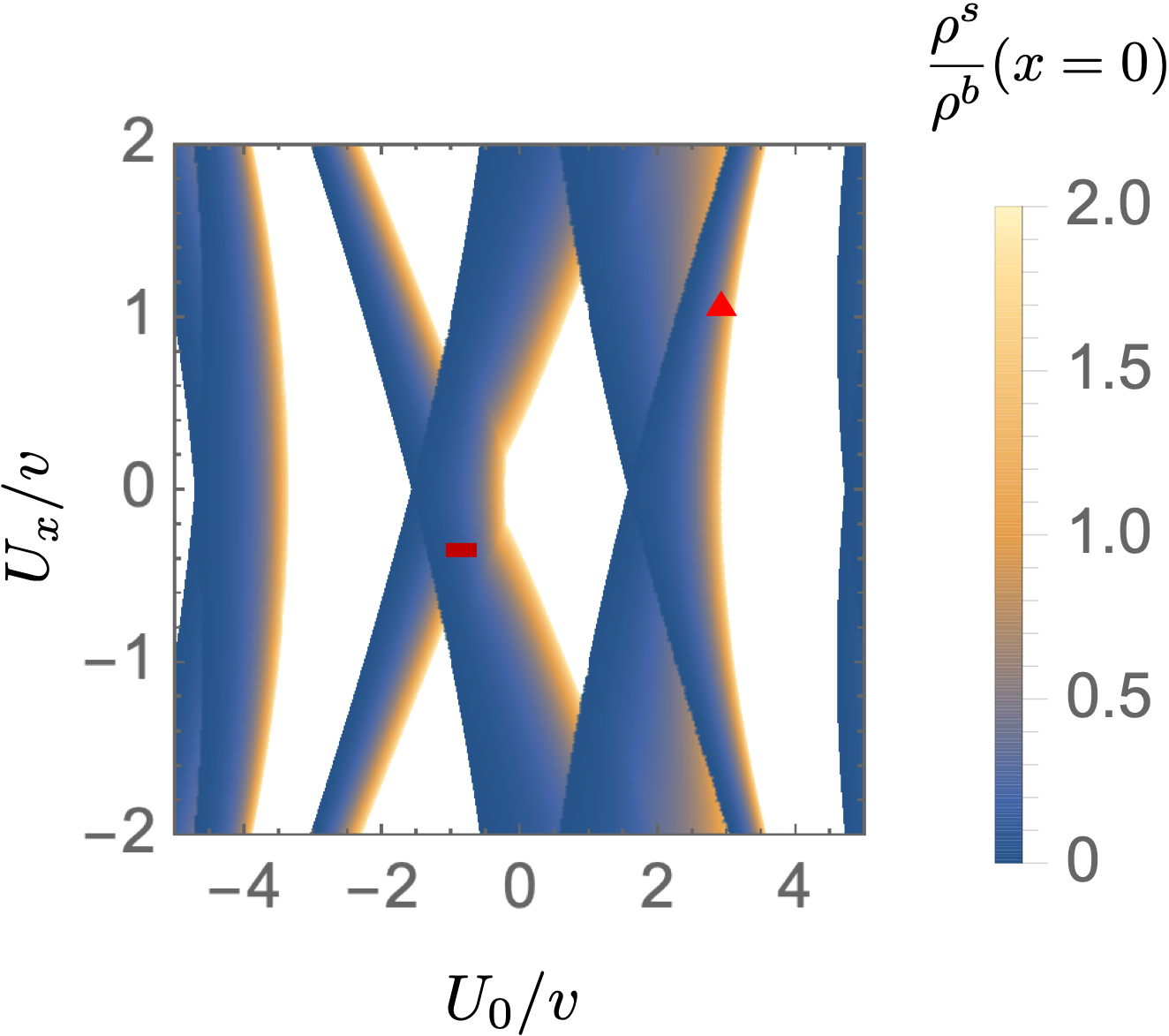}
\caption{Ratio of LDOS due to scattering and bound states at $x=0$ and at $\epsilon = 0.02$ when potentials $U_0$ and $U_x$ are varied. White space indicates absence of any bound states above Dirac point. $v=1$ and $E_G=1$.} \label{LDOS_at_0}
\end{figure}

\begin{figure}[H]
\includegraphics[width=0.5\textwidth]{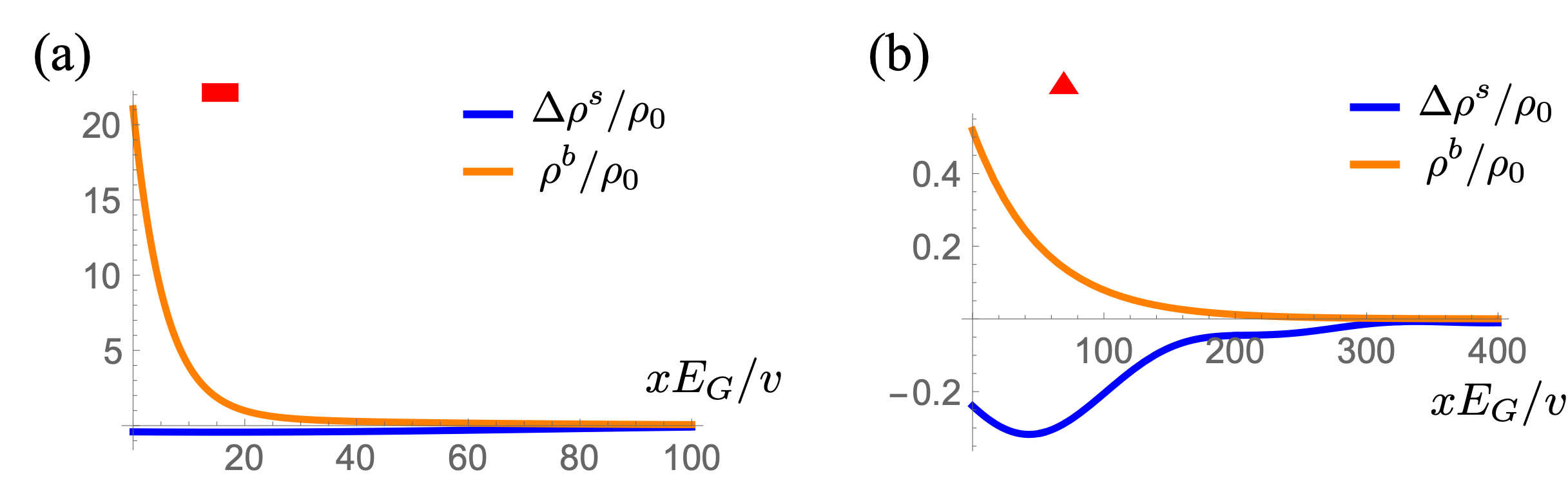}
\caption{Local Density of States at $\epsilon = 0.02$ when potentials are set to (a) $U_0/v = -1.0$, $U_x/v = -0.5$ and (b) $U_0/v = 3.0$, $U_x/v = 1.0$, and $\Delta \rho^s$ = $\rho_s-\rho_0$ where $\rho_0 = |\epsilon|/2\pi v^2$. The red rectangle and triangle indicate where in parameter space of Fig.~\ref{LDOS_at_0} $U_0$ and $U_x$ are located. $v=1$ and $E_G=1$.}
\label{LDOS}
\end{figure}

In figure \ref{LDOS_at_0}, we show the ratio of contribution to LDOS from scattering and bound states at $x=0$. When bound states exist, their contribution is comparable, or even dominant, over most of the parameter space. As an illustration, in figure \ref{LDOS}(a), we have chosen potentials such that bound state LDOS contribution dominate, while in figure \ref{LDOS}(b) we have chosen them to show comparable bound and scattering state contributions to LDOS.

\vspace{10mm}
\section{\label{app:spin_acc}Charge Current and Spin Accumulation Density}

{
In this appendix we compute the current density and spin accumulation for both bound and scattering states. In the limit $|U_x|/v\gg|U_0|/v\gg 1$ we obtain some exact results. }

 \subsection{Bound States}

\subsubsection{Spin accumulation $\mathcal{S}^b_x$ and current density $\mathcal{J}^b_y$}

As discussed in the main text, the in-plane spin accumulation and the current density have the same operator structure for the massless Dirac systems such as the one we consider.
Considering the bound state contribution, for the component of the spin (current) normal to (parallel to) the defect line we find

\begin{multline}
\mathcal{S}^{b\pm}_x(x) = \frac{\mathcal{J}^{b\pm}_y(x)}{ev} = \int_{-\Lambda}^{\mu} s_x^\pm(x,\epsilon) g^\pm(\epsilon) d\epsilon\\ \text{  or  } \int_{-\Lambda}^{\mu} \rho_{x}^{b\pm}
(\epsilon,x) d\epsilon,
\end{multline}
where $g(\epsilon) = \frac{1}{2\pi |v_\pm|}$ is the density of states for the bound state branches. Evaluating the energy integral, using Eq.~\eqref{rho_sx_U0Ux_bound}, we obtained



\begin{multline}
\mathcal{S}^{bu\pm}_x(x)=\frac{\mathcal{J}^{bu\pm}_y(x)}{ev} = \frac{1}{2\pi} \int_0^{\mu} \frac{\lambda_\pm^{(\epsilon)}}{v} e^{-2\lambda_\pm^{(\epsilon)}|x|} \text{sgn}(v_\pm) d\epsilon.
\end{multline}

Performing the integration we determine,

\begin{multline}
\mathcal{S}^{bu\pm}_x(x) = \frac{\mathcal{J}^{bu\pm}_y(x)}{ev} \\= \frac{1}{2\pi v} \text{sgn}(v_\pm) \mu\left(\frac{1-(1+2\lambda^{(\mu)}_\pm |x|)e^{-2\lambda^{(\mu)}_\pm|x|}}{4\lambda^{(\mu)}_\pm|x|^2}\right),
\label{spin_acc_bound_x_upper}
\end{multline}
where $\lambda^{(\mu)}_\pm = \frac{\mu}{v_\pm}\cos\alpha_\pm$. One can similarly perform the integral from $-\Lambda$ to $0$ and obtain,

\begin{multline}
\mathcal{S}^{bl\pm}_x(x) = \frac{\mathcal{J}^{bl\pm}_y(x)}{ev} \\= \frac{\Lambda}{2\pi v} \text{sgn}(v_\pm)\left(\frac{1-(1+2\lambda^{(-\Lambda)}_\pm |x|)e^{-2\lambda^{(-\Lambda)}_\pm|x|}}{4\lambda^{(-\Lambda)}_\pm|x|^2}\right).
\label{spin_acc_bound_x_lower}
\end{multline}

These give us the bound state contribution to in-plane spin accumulation and current density along $y$, $\mathcal{S}^{b\pm}_x(x)=\mathcal{S}^{bu\pm}_x(x)+\mathcal{S}^{bl\pm}_x(x)$ and $\mathcal{J}^{b\pm}_y(x)=\mathcal{J}^{bu\pm}_y(x)+\mathcal{J}^{bl\pm}_y(x)$, respectively.

\vspace{10mm}

We now compute the net bound state current by performing the spatial integral as well i.e. $J_y^b =e v\int_{-\Lambda}^\mu d\epsilon \int_{-\infty}^\infty dx \rho_x^b(\epsilon,x)$,


\begin{equation}
\frac{J_y^{b\pm}}{ev} = \frac{1}{2\pi v} \int_{-\Lambda}^\mu  \text{sgn}(v_{\pm}) \Theta(\lambda^\epsilon_\pm) d\epsilon.
\end{equation}

The total bound state current is $J_y^{b} = J_y^{b+}+ J_y^{b-}$. When $\mu>0$, the branches above Dirac point contribute $\text{sgn}(v_\pm)e\mu/2\pi$ and the branches below Dirac point contribute $\text{sgn}(v_\pm)e\Lambda/2\pi$ to the net current. When $\mu<0$, each branch below Dirac point contributes $\text{sgn}(v_\pm)e(\mu+\Lambda)/2\pi$.


\vspace{10mm}

\subsubsection{Spin accumulation $\mathcal{S}^b_z$}

The out-of-plane spin accumulation is given by

\begin{equation}
\mathcal{S}^{b\pm}_z(x) = \int_{-\Lambda}^{\mu} s_z^\pm(x,\epsilon) g^\pm(\epsilon) d\epsilon \text{  or  } \int_{-\Lambda}^{\mu} \rho_{z}^{b\pm}
(\epsilon,x) d\epsilon,
\end{equation}
where, once again,  $g(\epsilon) = \frac{1}{2\pi |v_\pm|}$ is the density of states for the bound state branches. Evaluating the energy integral, using Eq.~\eqref{rho_sz_U0Ux_bound}, we obtained

\begin{multline}
\mathcal{S}^{bu\pm}_z(x) = \int_{0}^\mu -\frac{1}{2\pi}\frac{(\lambda_\pm^{(\epsilon)})^2}{\epsilon} e^{-2\lambda_\pm^{(\epsilon)}|x|} \text{sgn}\left(x{v_\pm}\right)  d\epsilon.
\end{multline}

Performing the integral, we find
\begin{multline}
\mathcal{S}^{{bu\pm}}_z(x) \\= -\frac{\mu}{2\pi}  \text{sgn}\left({x}\right) \frac{\cos\alpha_\pm}{|v_\pm|} \left(\frac{1-(1+2\lambda^{(\mu)}_\pm |x|)e^{-2\lambda^{(\mu)}_\pm|x|}}{4\lambda^{(\mu)}_\pm|x|^2}\right)
\label{spin_acc_bound_z_upper}
\end{multline}

Similarly performing the energy integral from $-\Lambda$ to $0$, we obtain

\begin{multline}
\mathcal{S}^{bl\pm}_z(x) \\= -\frac{\Lambda}{2\pi}  \text{sgn}\left({x}\right) \frac{\cos\alpha_\pm}{|v_\pm|} \left(\frac{1-(1+2\lambda^{(-\Lambda)}_\pm |x|)e^{-2\lambda^{(-\Lambda)}_\pm|x|}}{4\lambda^{(-\Lambda)}_\pm|x|^2}\right).
\label{spin_acc_bound_z_lower}
\end{multline}

These give us the bound state contribution to out-of-plane spin accumulation, $\mathcal{S}^{b\pm}_z(x)=\mathcal{S}^{bu\pm}_z(x)+\mathcal{S}^{bl\pm}_z(x)$.

%
%
%
%

\subsection{Scattering States}

\subsubsection{Spin accumulation $\mathcal{S}^s_{x}$ and current density $\mathcal{J}^s_{y}$}
\vspace{5mm}

Current density and in-plane spin accumulation is given by

\begin{multline}
\mathcal{S}^s_{x}(x) =\frac{\mathcal{J}^s_y (x)}{ev} = \int_{-\Lambda}^\mu \rho^s_{x} (\epsilon,x) d\epsilon \\ =  \int_{0}^\mu \rho^s_{x} (\epsilon,x) d\epsilon - \int^{-\Lambda}_0 \rho^s_{x} (\epsilon,x) d\epsilon
\\ = \mathcal{S}_{x,z}^{s,\mu}	-\mathcal{S}_{x,z}^{s,-\Lambda}
\label{sx_density_def}
\end{multline}

In the limit $\mu|x|/v>>1$, the dominant contribution is from angles near $\theta=0$, thus using stationary phase approximation~\cite{,Bhattacharya1979} we obtain,

{
\begin{multline}
\mathcal{S}_x^{s,\mu}(x) = \int_{0}^\mu \rho^s_{x} (\epsilon,x) d\epsilon \\= \left(\frac{\mu}{v}\right)^2K_x^\mu[U_0,U_x] \frac{\cos(2|\mu x|/v + \tilde\phi_x)}{(|\mu x|/v)^{3/2}}
\end{multline}
where

\begin{align}
K_x^\mu [U_0,U_x] = \text{sgn}(\mu)\frac{\sqrt{\pi}}{(2\pi)^2
} \sqrt{C_4^2+S_4^2}, {\hspace{3mm}} \\ \tilde\phi_x = -\frac{\pi}{4} - \tan^{-1}\left(\frac{S_4}{C_4}\right),
\end{align}
}
%

\begin{equation}
C_4 = -\text{sgn}(\mu)\frac{\eta  {U_x} \sinh\eta  \cosh\eta }{\eta ^2 \cosh ^2\eta +{U_0}^2 \sinh ^2\eta },
\end{equation}
and

\begin{equation}
S_4 = -\frac{{U_0} {U_x} \sinh^2\eta }{\eta ^2 \cosh^2\eta +{U_0}^2 \sinh ^2\eta }.
\end{equation}

%
%
%
%

Thus, the oscillations have a period $ \pi v/|\mu| \sim \pi/k_F$, and oscillation amplitude decays as 
{$\sim \sqrt{|\mu|/v}/|x|^{3/2}$}. Second integral in Eq.~\eqref{sx_density_def} is obtained by substituting $\mu\rightarrow -\Lambda$ above. The Freidel oscillations contribution, thus obtained, oscillate with a period $\pi v/|\Lambda|$ while amplitude of oscillations decays as {$\sim \sqrt{|\Lambda|/v}/|x|^{3/2}$}. For $|\Lambda|>>|\mu|$, the total contribution oscillates with period $\pi v/|\Lambda|$. For $|x|>>v/|\mu|$ the ratio of the two contributions $\sim\sqrt{|\Lambda/\mu|}$. 


\vspace{5mm}

In the limit, $U_x>>U_0$ and $U_x>>1$, we  obtain the current contribution from scattering states by performing spatial integration and integration over energy on $\rho^s_x$ in Eq.~\eqref{LDOS_x_large_Ux}. This gives

%

\begin{equation}
\frac{J_y^s}{ev}=\int_{-\Lambda}^\mu d\epsilon \int_{-\infty}^\infty dx \rho_x^s(\epsilon,x) = \text{sgn}(U_x)\frac{(|\mu| -\Lambda)}{2\pi v}.
\end{equation}

\vspace{5mm}

\subsubsection{Spin accumulation $\mathcal{S}^s_{z}$}

Out-of-plane spin accumulation is given by

\begin{equation}
\mathcal{S}^s_{z}(x) = \int_{-\Lambda}^\mu \rho^s_{z} (\epsilon,x)d\epsilon =  \int_{0}^\mu \rho^s_{x} (\epsilon,x) d\epsilon - \int^{-\Lambda}_0 \rho^s_{x} (\epsilon,x) d\epsilon
\label{sz_density_def}
\end{equation}

Again, as above, in the limit $\mu|x|/v>>1$, the dominant contribution is from angles near $\theta=0$, and we use the stationary phase approximation~\cite{Villain2016,callaway1976book,Bhattacharya1979} to obtain,

{

\begin{equation}
\mathcal{S}_z^{s,\mu}(x) = \left(\frac{\mu}{v}\right)^2 K_z^\mu[U_0,U_x] \frac{\cos(2|\mu x|/v + \tilde\phi_z)}{(|\mu x|/v)^{3/2}}
\end{equation}

where

\begin{align}
K_z^\mu[U_0,U_x] = \frac{\text{sgn}(x)}{(2\pi)^2} \sqrt{\pi}\sqrt{C_5^2+S_5^2},\\
\tilde{\phi_z} = -\frac{\pi}{4} - \tan^{-1}\left(\frac{S_5}{C_5}\right),
\end{align}
}

\begin{equation}
C_5 = \text{sgn}(\mu)\frac{{U_0} {U_x} \sinh ^2\eta  }{\eta^2 \cosh^2\eta +{U_0}^2 \sinh^2\eta },
\end{equation}
and

\begin{equation}
S_5 = -\frac{\eta  {U_x} \sinh\eta \cosh\eta}{\eta^2 \cosh^2\eta +{U_0}^2 \sinh^2\eta }.
\end{equation}



{Again, the second integral in Eq.~\eqref{sz_density_def} is obtained by substituting $\mu\rightarrow -\Lambda$ above. The oscillations in $\mathcal{S}_z^{s,\beta}(x)$ have the same periods and amplitude decay as oscillations in $\mathcal{S}_x^{s,\beta}(x)$. }




%

\vspace{10mm}

\section{\label{sec:born_app}Scattering from impurities}

In this appendix, we calculate the broadening of bound state dispersion when electrons scatter into dispersive states using Eq.~\eqref{broadening_def}. 
We first calculate $\sum_{\vec{k}} |\braket{\psi_{s}|\hat{V}|\psi_b}|^2\delta(\epsilon-E(\bm{k}))$ for a single impurity located at $\vec{R_0} = (x_0,y_0)$ and then average over the impurity position to get the result for random distribution of point-like impurities. The potential due to the impurity is given by $V_0\delta(\vec{r}-\vec{R}_0)$. Without loss of generality, we can assume that $x_0 > 0$. The scattering state wavefunction of an electron coming in from $x\rightarrow \infty$, $\ket{\psi_2^h(x>0)} = \ket{\psi^{h }_{2,i}} + \ket{\psi^{h}_{2,\mathcal{R}}}$ is known from Eq.~\eqref{wavefn_scatt_2_right}. Similarly, a particle coming in from $x\rightarrow -\infty$ will have wavefunction $\ket{\psi_1^h (x>0)} = \ket{\psi^{h}_{1,\mathcal{T}}}$ We evaluate $|\braket{\psi_{s}|\hat{V}|\psi_b}|^2$  for $E>0$ but suppress index $h=+1$ below for brevity 

\begin{widetext}
\begin{multline}
|\braket{\psi_{2}^i|V_0\delta(\vec{r}-\vec{R}_0)|\psi_b} + \braket{\psi_{2}^\mathcal{R}|V_0\delta(\vec{r}-\vec{R}_0)|\psi_b}|^2 + |\braket{\psi^\mathcal{T}_1|V_0\delta(\vec{r}-\vec{R}_0)|\psi_b} |^2 = \left(\lim_{a,b\rightarrow \infty}\frac{V_0^2}{4Ab} \right) \lambda_+ e^{-2\lambda_+ x_0} \\\Bigg[2\left( \chi_+^2 + \chi_-^2 + 2 \chi_+ \chi_- \sin\theta \right) + 2 \text{Re}\Bigg[\mathcal{R}_2^{*} e^{-2ik_x x_0} \Bigg(\chi_-^2 - \chi_+^2 e^{-2i\theta} +2i\chi_+ \chi_- e^{-i\theta}\Bigg)\Bigg]\Bigg].
\label{overlap}
\end{multline}
\end{widetext}
where $\chi_\pm = \sqrt{1\pm\lambda_+/k_y'} = \sqrt{1\pm\cos\alpha_+}$ and $\mathcal{R}_2$ is given by $h=+1$ expression in Eq.~\eqref{r2_U0Ux}. Now we average the above result to get the broadening for a uniform distribution of impurities.



\begin{figure}[t]
\includegraphics[width=0.5\textwidth]{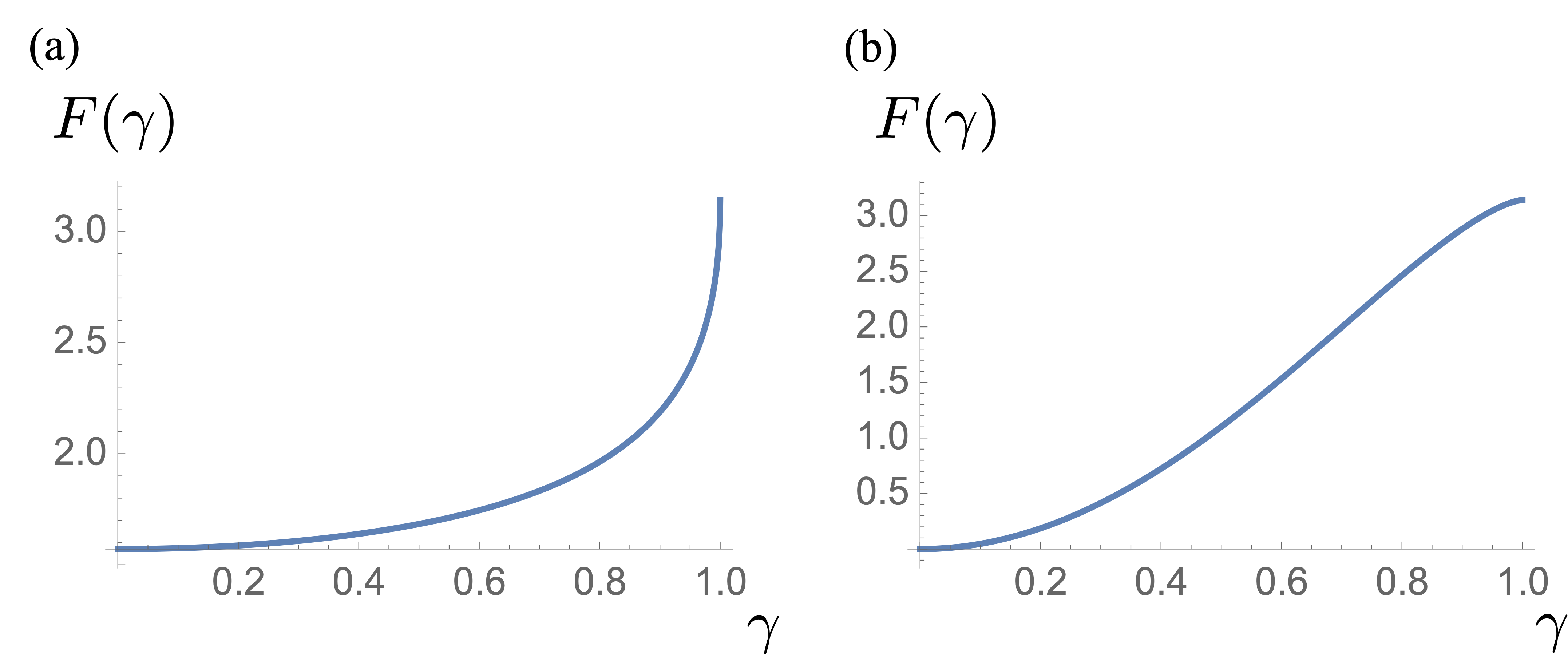}
\caption{Spinor overlap $F(\gamma)$ for (a) electrostatic potential and (b) magnetic potential at the defect when ratio $\gamma = v_+/v$ is varied from flat state ($v_+=0$) to close to Dirac velocity $v_+\approx v$.}
\label{F}
\end{figure}

\begin{multline}
|\braket{\psi_s|V|\psi_b}|^2 = \frac{2}{ab}\int_{0}^{a/2}dx_0\int_{-b/2}^{b/2}dy_0 \\
|\braket{\psi^i_2|V_0\delta(\vec{r}-\vec{R}_0)|\psi_b} + \braket{\psi^\mathcal{R}_2|V_0\delta(\vec{r}-\vec{R}_0)|\psi_b}|^2 \\+ |\braket{\psi^\mathcal{T}_1|V_0\delta(\vec{r}-\vec{R}_0)|\psi_b} |^2.
\end{multline}

%
%


%
%
%
%

Broadening of bound state dispersion is given by


\begin{multline}
\Gamma = \left(\lim_{a, b \rightarrow \infty} \frac{V_0^2  }{A}\right) \frac{\epsilon}{2\pi v^2}  \\ \int_{-\pi/2}^{\pi/2} d\theta  \frac{1}{2} \Bigg[\left( \chi_+^2 + \chi_-^2 + 2 \chi_+ \chi_- \sin\theta \right) \\+ \text{Re}\Bigg[\mathcal{R}_2^{*} \frac{\lambda_+}{\lambda_+ + ik_x} \Bigg(\chi_-^2 - \chi_+^2 e^{-2i\theta} +2i\chi_+ \chi_- e^{-i\theta}\Bigg)\Bigg]\Bigg].
\end{multline}

The expression has the form

\begin{equation}
\Gamma = n_{imp} V_0^2 \rho_0(\epsilon) F(\gamma),
\end{equation}
where $n_{imp}$ is the impurity concentration, $\rho_0(\epsilon) = \frac{\epsilon}{2\pi v^2}$ and $F(\gamma)$ the spinor overlap given by

\begin{multline}
F(\gamma) = \int_{-\pi/2}^{\pi/2} d\theta  \frac{1}{2} \Bigg[\left( \chi_+^2 + \chi_-^2 + 2 \chi_+ \chi_- \sin\theta \right) \\+ \text{Re}\Bigg[\mathcal{R}_2^{*} \frac{\lambda_+}{\lambda_+ + ik_x} \Bigg(\chi_-^2 - \chi_+^2 e^{-2i\theta} +2i\chi_+ \chi_- e^{-i\theta}\Bigg)\Bigg]\Bigg].
\label{F_eq}
\end{multline}

In figure \ref{F}(a) and \ref{F}(b), we plot the $F(\gamma)$ in presence of only electrostatic and magnetic scattering respectively. Note that $0\leq F(\gamma)\leq \pi$, and hence this function simply gives a prefactor to the characteristic broadening described in the main text.   

\end{document}